\documentclass[12pt,useAMS,usenatbib]{mn2e}
\usepackage{graphicx}
\usepackage{txfonts}
\usepackage{rotating}
\usepackage{lscape}
\usepackage{longtable}
\title[Hard X-ray spectra of AGN in the INTEGRAL complete sample]
 {Hard X-ray spectra of AGN in the INTEGRAL complete sample}
\author[M. Molina et al.]
{M. Molina$^1$, L. Bassani$^1$, A. Malizia$^1$, J.B. Stephen$^1$, A.J. Bird$^2$, A. Bazzano$^3$, P. Ubertini$^3$\\
$^1$IASF/INAF, via Gobetti 101, I-40129 Bologna, Italy\\
$^2$School of Physics and Astronomy, University of Southampton,
        SO17 1BJ, Southampton, U.K., \\
$^3$IAPS/INAF, Via del Fosso del Cavaliere 100, 00133 Rome, Italy}

\begin{document}

\date{}

\pagerange{\pageref{firstpage}--\pageref{lastpage}} \pubyear{2009}

\maketitle

\label{firstpage}
        
\begin{abstract}
In this paper, we present the hard X-ray spectral analysis of
a complete sample of AGN detected by \emph{INTEGRAL/IBIS}. In conjunction with \emph{IBIS} spectra, 
we make use of \emph{Swift/BAT} data, 
with the aim of cross-calibrating the two instruments, studying source variability 
and constraining some important spectral parameters. We find that flux variability is
present in at least 14\% of the sample, while spectral variability is found only in one object.
There is general good agreement between \emph{BAT} and
\emph{IBIS} spectra, despite a systematic mismatch of about 22\% in normalisation.
When fitted with a simple power-law model, type 1 and type 2 sources appear to have very similar
average photon indices, suggesting that they are powered by the same mechanism.
As expected, we also find that a simple power-law does not always describe the data sufficiently well, thus
indicating a certain degree of spectral complexity, which can be ascribed to features
like a high energy cut-off and/or a reflection component. Fixing the reflection to be 0, 1 or 2, we find that 
our sample covers quite a large 
range in photon indices as well as cut-off energies; however, the spread is due only to a small 
number of objects, while the majority of the AGN lie within well defined boundaries of photon index 
(1$\le$$\Gamma$$\le$2) and cut-off energy (30$\le$E$_{\rm cut}$$\le$300\,keV).
\end{abstract}


\section{Introduction}

The study of active galactic nuclei (AGN) above 10\,keV is essential if one 
wants to study non-thermal processes and observe those sources
which are strongly affected by absorption in the soft X-ray band. 
Another advantage is the possibility of having information on spectral features such as the high energy cut-off
and the reflection fraction, which cannot be explored with observations performed below 10\,keV.
The determination of these parameters is important for many reasons: they  
provide an insight into the physical properties
of the region around the central power source, they play a key role in synthesis models  
of the cosmic X-ray background and are important ingredients 
for unification theories and torus studies (e.g \citealt{Urry:1995, Elitzur:2006, elitzur12}). 
Although high energy measurements of AGN have been made in the past, mainly with the \emph{BeppoSAX}
satellite (e.g. \citealt{Perola:2002}; \citealt{Dadina:2007}), 
these did not generally pertain to a large and complete sample of sources and were limited to a few bright
nearby objects. 
This trend is now changing since the launch of facilities like \emph{INTEGRAL}
\citep{winkler03} and \emph{Swift} \citep{gehrels04}, both
having imaging and spectroscopy capabilities on a large field of view.
Indeed, both \emph{IBIS} \citep{Ubertini:2003} on board \emph{INTEGRAL}
and \emph{BAT} \citep{barthelmy05} on board \emph{Swift} have been surveying
the high energy sky for many years now, providing many detections of both new and previously known active galaxies, 
which can now be studied in a systematic way for the first time above 10\,keV.

Cross-calibration between \emph{BAT} and \emph{IBIS} is important if one wants
to enhance the signal-to-noise ratio or to study source variability over long timescales.
Once calibrated, data from both instruments can be used alone or in conjunction with low
energy observations to obtain broadband information and to put stronger constraints on spectral parameters.

In this paper, we present the hard X-ray spectral analysis,
conducted in the 17--150\,keV energy range, of a complete
sample of 87 AGN detected by \emph{INTEGRAL/IBIS} \citep{Malizia:2009}.
For 80 of these sources we also used \emph{BAT} data, with the aim
of cross-calibrating the two instruments, studying source variability 
and constraining the primary continuum.   
This work can also be useful for future exploitation of the \emph{IBIS/BAT}
AGN archive in conjunction with low energy instrument data, 
for example as an aid to \emph{NuSTAR} observations.

\section{Complete Sample and Data Analysis}

The \emph{INTEGRAL} complete sample of AGN has been extracted 
from a set of sources listed in the third \emph{IBIS}
survey \citep{Bird:2007} and having a firm identification with an AGN. 
This complete sample consists of 87 active galaxies of various optical classifications:
41 type 1 AGN (Seyfert 1, 1.2 and 1.5), 33 type 2 AGN (Seyfert 1.9
and 2), 5 narrow line Seyfert 1s (NLSy1s) and 8 Blazars (QSOs and BL Lacs).
Full details on the extraction of the complete sample are given in \citet{Malizia:2009}.
Table~\ref{sample1} lists relevant information (source name, 
Galactic column density, redshift and AGN class) for all objects in the sample.
The \emph{INTEGRAL/IBIS} data reported here consist of several pointings  
performed in the period comprised between launch (October 2002) and the end of April
2008, for a total of almost 40000 science windows. 
\emph{IBIS/ISGRI} images for each available pointing were generated 
in various energy bands using the ISDC offline scientific 
analysis software OSA \citep{Goldwurm:2003} version 7.0. 
Count rates at the position of the source were extracted 
from individual images using data from the \emph{ISGRI} 
detector \citep{Lebrun:2003} in order to provide light curves in various energy bands; from these 
light curves, average fluxes were then extracted and combined to produce an average, 
14 channels (covering the 17\,keV to 300\,keV energy range), source spectrum 
(see \citealt{Bird:2007} and \citealt{Bird:2010} for details). This 
spectral extraction has been adopted to deal with the large number of objects listed in the survey. 
This method has also been fully tested over the 20--100\,keV band in a number of studies of 
AGN spectral properties (e.g. \citealt{Molina:2009}). 
\emph{Swift/BAT} spectra were retrieved on the web\footnote{http://swift.gsfc.nasa.gov/docs/swift/results/bs58mon/}; these 
spectra are from the first 58 months of operations of the
\emph{Swift/BAT} telescope \citep{baumgartner10} and provide information over the 14--195\,keV band
(note that the spectra have an 8 channel binning).
In order to overlap with \emph{IBIS} data and cross-calibrate the two instruments,
we explore here a slightly narrower energy band (17--150\,keV).

\emph{BAT} and \emph{IBIS} spectra have been fitted
with \texttt{XSPEC} version 12.7.1, using the $\chi^2$ statistics, 
assuming that the data are Gaussian; in the following 
analysis, each parameter error is quoted at 90\% confidence level ($\Delta\chi^2$=2.71). 

\begin{table*}
\footnotesize
\centering
\caption{{\bf The \emph{INTEGRAL/IBIS} Complete Sample}}
\label{sample1}
\vspace{0.2cm}
\begin{tabular}{lccr|lccr}
\hline
{\bf Name}&{\bf N$_{\bf \rm H}^{\rm \bf gal}$}&{\bf z}& {\bf type}& {\bf Name}&{\bf N$_{\bf \rm H}^{\rm \bf gal}$}&{\bf z}&{\bf type}\\
          &{\bf 10$^{\bf 22}$ cm$^{-2}$}      &                     &    &          &{\bf 10$^{\bf 22}$ cm$^{-2}$}      &                     &        \\ 
\hline
IGR J00333+6122    & 0.55 & 0.1050 & Sy 1.5&4U 1344-60	   & 1.07 & 0.0130 & Sy 1.5\\ 
1ES 0033+595       & 0.04 & 0.0860 &BL Lac &IC 4329A           & 0.04 & 0.0160 & Sy 1.2\\
NGC 788            & 0.02 & 0.0136 & Sy 2  &Circinus Galaxy    & 0.56 & 0.0014 & Sy 2  \\
NGC 1068           & 0.03 & 0.0038 & Sy 2  &NGC 5506           & 0.04 & 0.0062 & Sy 1.9\\
QSO B0241+62       & 0.75 & 0.0440 & Sy 1  &ESO 511-G030       & 0.05 & 0.0224 & Sy 1  \\       
NGC 1142           & 0.06 & 0.0288 & Sy 2  &IGR J14515-5542    & 0.53 & 0.0180 & Sy 2  \\
B3 B0309+411B      & 0.13 & 0.1360 & Sy 1  &IC 4518A           & 0.09 & 0.0163 & Sy 2  \\
NGC 1275           & 0.15 & 0.0175 & Sy 2  &IGR J16024-6107    & 0.29 & 0.0110 & Sy 2  \\
3C 111 	           & 0.32 & 0.0485 & Sy 1  &IGR J16119-6036    & 0.23 & 0.0160 & Sy 1  \\
LEDA 168563	       & 0.54 & 0.0290 & Sy 1  &IGR J16185-5928    & 0.25 & 0.0350 & NLSy1 \\
4U 0517+17	       & 0.22 & 0.0179 & Sy 1.5&IGR J16351-5806    & 0.25 & 0.0091 & Sy 2  \\
MCG+08-11-011      & 0.20 & 0.0205 & Sy 1.5&IGR J16385-2057    & 0.12 & 0.0269 & NLSy1 \\
Mkn 3              & 0.08 & 0.0135 & Sy 2  &IGR J16426+6536    & 0.03 & 0.3230 & NLSy1 \\
Mrk 6	           & 0.06 & 0.0188 & Sy 1.5&IGR J16482-3036    & 0.18 & 0.0310 & Sy 1  \\
IGR J07565-4139    & 0.47 & 0.0210 & Sy 2  &IGR J16558-5203    & 0.30 & 0.0540 & Sy 1.2\\
IGR J07597-3842	   & 0.60 & 0.0400 & Sy 1.2&Swift J1656.3-3302 & 0.22 & 2.4000 &Blazar \\
ESO 209-12         & 0.19 & 0.0396 & Sy 1.5&NGC 6300           & 0.09 & 0.0037 & Sy 2  \\
QSO B0836+710      & 0.03 & 2.1720 &Blazar &GRS 1734-292       & 0.77 & 0.0214 & Sy 1  \\
FRL 1146	           & 0.40 & 0.0316 & Sy 1.5&2E 1739.1-1210     & 0.21 & 0.0370 & Sy 1  \\
Swift J0917.2-6221 & 0.19 & 0.0573 & Sy 1  &IGR J17488-3253    & 0.53 & 0.0200 & Sy 1  \\
MCG-05-23-16       & 0.08 & 0.0085 & Sy 2  &IGR J17513-2011    & 0.35 & 0.0470 & Sy 1.9\\
IGR J09523-6231    & 0.27 & 0.2520 & Sy 1.9&IGR J18027-1455    & 0.50 & 0.0350 & Sy 1  \\
Swift J1009.3-4250 & 0.11 & 0.0330 & Sy 2  &IGR J18249-3243    & 0.12 & 0.3550 & Sy 1  \\
NGC 3281           & 0.06 & 0.0115 & Sy 2  &IGR J18259-0706    & 0.62 & 0.0370 & Sy 1  \\
Swift J1038.8-4942 & 0.23 & 0.0600 & Sy 1.5&PKS 1830-211       & 0.22 & 2.5070 &Blazar \\
IGR J10404-4625    & 0.14 & 0.2392 & Sy 2  &ESO 103-35         & 0.08 & 0.0133 & Sy 2  \\
NGC 3783	           & 0.08 & 0.0097 & Sy 1  &3C 390.3           & 0.04 & 0.0561 & Sy 1  \\
IGR J12026-5349    & 0.16 & 0.0280 & Sy 2  &2E 1853.7+1534     & 0.39 & 0.0840 & Sy 1  \\
NGC 4151	           & 0.02 & 0.0033 & Sy 1.5&IGR J19378-0617    & 0.15 & 0.0106 & NLSy1 \\
4C 04.42           & 0.02 & 0.9650 & QSO   &NGC 6814	   & 0.13 & 0.0052 & Sy 1.5\\
Mrk 50	           & 0.02 & 0.0234 & Sy 1  &Cyg A              & 0.35 & 0.0561 & Sy 2  \\
NGC 4388           & 0.03 & 0.0084 & Sy 2  &IGR J20186+4043    & 1.20 & 0.0144 & Sy 2  \\
3C 273             & 0.02 & 0.1583 & QSO   &4C 74.26           & 0.12 & 0.1040 & Sy 1  \\
NGC 4507           & 0.07 & 0.0118 & Sy 2  &S5 2116+81         & 0.07 & 0.0840 & Sy 1  \\
LEDA 170194        & 0.04 & 0.0360 & Sy 2  &IGR J21247+5058    & 1.11 & 0.0200 & Sy 1  \\
NGC 4593	           & 0.02 & 0.0090 & Sy 1  &Swift J2127.4+5654 & 0.79 & 0.0140 & NLSy1 \\
IGR J12415-5750    & 0.30 & 0.0244 & Sy 1  &RX J2135+4728      & 0.32 & 0.0250 & Sy 1  \\
3C 279             & 0.02 & 0.5362 &Blazar &NGC 7172           & 0.02 & 0.0087 & Sy 2  \\
NGC 4945           & 0.16 & 0.0019 & Sy 2  &BL Lac             & 0.21 & 0.0686 &BL Lac \\
IGR J13091+1137    & 0.02 & 0.0251 & Sy 2  &MR 2251-178        & 0.03 & 0.0640 & Sy 1  \\
IGR J13109-5552    & 0.22 & 0.0850 & Sy 1  &MCG-02-58-022      & 0.04 & 0.0469 & Sy 1.5\\
Cen A              & 0.09 & 0.0018 & Sy 2  &IGR J23308+7120    & 0.30 & 0.0370 & Sy 2  \\
MCG-06-30-015      & 0.04 & 0.0077 & Sy 1.2&IGR J23524+5842    & 0.57 & 0.1640 & Sy 2  \\
NGC 5252           & 0.02 & 0.0230 & Sy 2  &                   &      &        &       \\  
\hline
\end{tabular}
\end{table*}

\section{\emph{INTEGRAL/IBIS} versus \emph{Swift/BAT}: source variability}\label{var}

X-ray variability, both in flux and in spectral shape,
is one of the most useful tools to probe
the physical processes taking place in the surroundings
of the central black hole, being related to the physical size and state of the X-ray 
emitting region. While AGN variability in the 2--10\,keV energy range has been well
studied, changes above 10\,keV have been much less sampled.
In the past, both \emph{CGRO/OSSE} and \emph{BeppoSAX/PDS}
have conducted studies on AGN variability in the
hard X-ray band, but none of these instruments had the
possibility of performing the analysis on a complete sample of AGN.
This is now possible using the present dataset: indeed, 80 out of 87
AGN listed in Table~\ref{sample1} have both \emph{IBIS} and
\emph{BAT} data available, making average flux/shape
comparison, instrument calibration and spectral studies possible.
 
Despite being averaged over a long period, the two datasets
have been derived from observations taken over different 
time intervals and so can be used to search for source variability, if present.
The \emph{IBIS} and \emph{BAT} spectra have therefore been fitted
individually, employing a simple power-law absorbed by intrinsic 
column density.
Although the introduction of the intrinsic column density in the 
model might not greatly affect sources with small N$_{\rm H}$,
heavy absorption as measured, for example, in Compton Thick (CT) objects, can 
have nonetheless an effect above 10\,keV.
Tables \ref{ind_fit_type1}, \ref{ind_fit_type2} and \ref{ind_fit_qso} report
the fit results for each class of objects and for both instruments, i.e. photon index, 
20--100\,keV flux, $\chi^2$ and degrees of freedom; 
N$_{\rm H}$ values are those reported
in \citet{Malizia:2009}.

\begin{figure}
\centering
\includegraphics[width=1.05\linewidth]{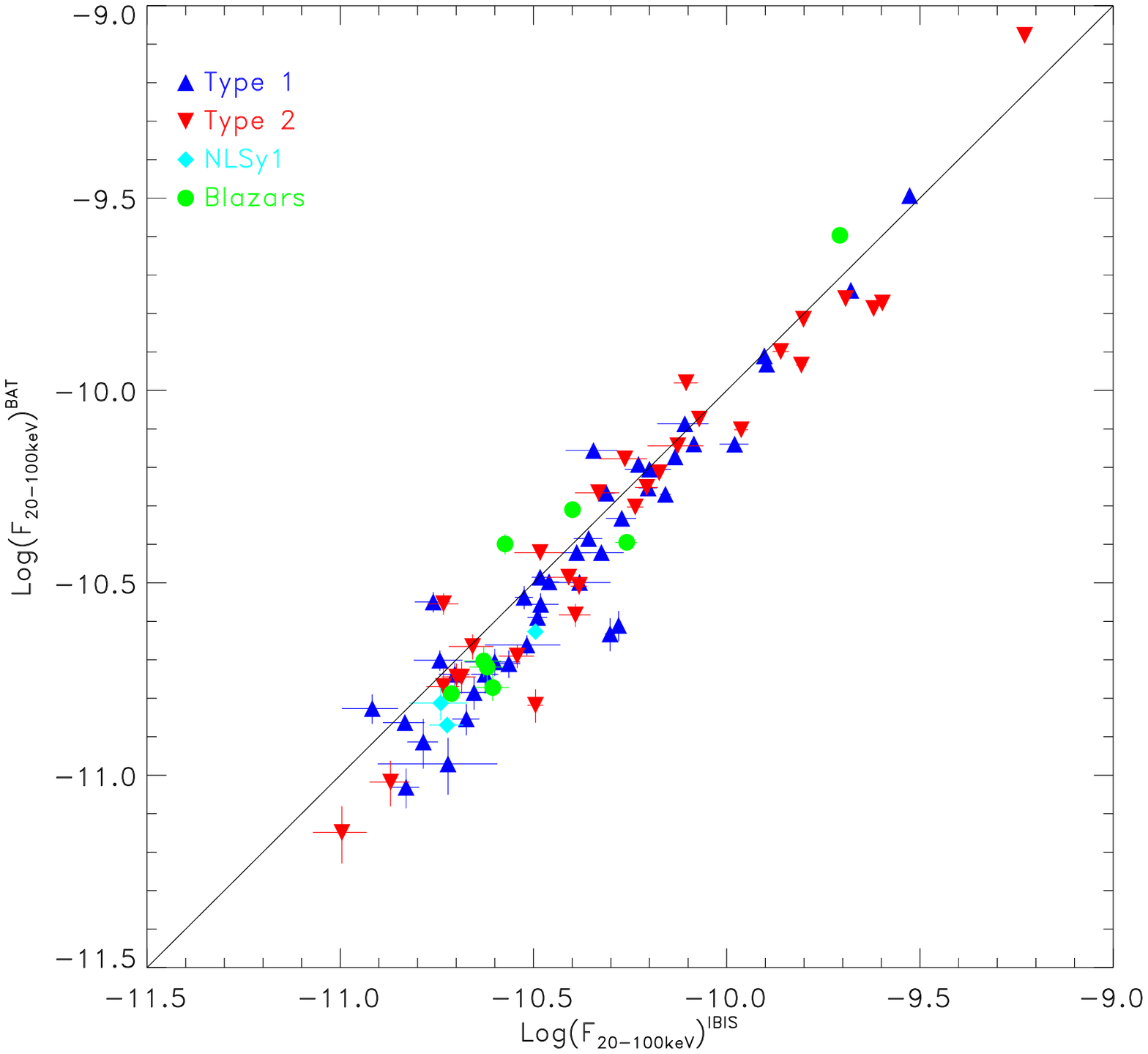}
\caption{20-100 keV fluxes derived from individual fits of the 80 sources for
which \emph{IBIS} and \emph{BAT} spectra are available. Blue up-side triangles represent type 1 AGN, red down-side
triangles represent type 2 AGN, cyan diamonds are Narrow Line Seyfert 1s, while green filled circles
represent Blazars.}
\label{flux_bat_ibis}
\end{figure}

Figure \ref{flux_bat_ibis} shows the 20--100\,keV \emph{IBIS}
versus \emph{BAT} fluxes, both obtained from
the individual power-law fits to the high energy spectra. It is quite evident from the
plot that there are significant deviations from the 1:1 line, 
suggesting that above 10\,keV long term variability is likely 
and is detected in all types of objects.
Also evident from the figure is the tendency for many points
to be located below the 1:1 line, especially for dimmer sources:
this can be interpreted as a sign of some systematic differences between the two instruments.

To quantify this mismatch in normalisation, we have calculated the ratios F$_{\rm IBIS}$/F$_{\rm BAT}$
for the entire sample and fit a gaussian to their distribution in order to 
estimate the mean and the relative error: we obtain a mean of 1.22 and an error of 0.036. 
We have then multiplied the \emph{BAT} flux by this mean value,
taking into account the errors and then calculated again
F$_{\rm IBIS}$/F$_{\rm BAT}$ to ensure that the new distribution is compatible 
within errors with a gaussian having mean 1 (the new value of the mean is 1.03
with an error of 0.03).
In the following considerations, this corrected F$_{\rm BAT}$ value will be considered. 

To evaluate the presence of flux variability, we then estimated 
F$_{\rm IBIS}$-F$_{\rm BAT}$/$\sqrt{(\sigma_{\rm IBIS}^2+\sigma_{\rm BAT}^2)}$, made the histogram 
of the values obtained and fitted a gaussian to it; 
at this point we estimated how many sources deviate from this 
gaussian significantly and are therefore likely to be variable.
In particular, we consider a source to be variable if its 20--100\,keV flux
has changed by more than 3$\sigma$ between the two instruments flux averages.  
In this way, we find that 11 sources (see last column in Tables~\ref{ind_fit_type1},  
\ref{ind_fit_type2} and \ref{ind_fit_qso})
out of 80 (around 14\% of the sample) have undergone a change in their flux. 
In particular, 4 of these variable AGN are type 1 sources,
2 are Blazars and 5 are type 2 objects. 
The amount of variability observed ranges from 0.5 to about a factor of 2. 
We emphasise again that these values are corrected for the systematic
mismatch in normalization between the two instruments 
discussed above. 

\begin{figure}
\centering
\includegraphics[width=1.05\linewidth]{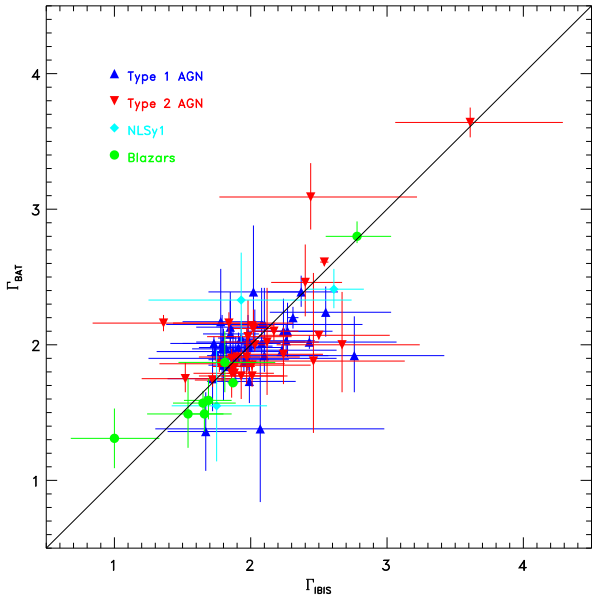}\\
\caption{Simple power-law photon indices, derived from individual
fits of \emph{BAT} and \emph{IBIS} spectra. Only sources with both datasets (80 out of 87) are shown.
Symbols and colours are as in Figure~\ref{flux_bat_ibis}.}
\label{gamma_bat_ibis}
\end{figure}

AGN flux variability at hard X-rays has been discussed by 
\citet{Beckmann:2007} using \emph{Swift/BAT} data for 
44 bright AGN: they found that Blazars show stronger variability 
than type 1 and type 2 AGN, which instead exhibit 10\% flux variations or 
more in at least one third of the sources analysed.
The comparison between our results with those of \citet{Beckmann:2007} 
is difficult, due to the different methods of analysis employed. 
We do not see more variability in Blazars than in Seyferts, 
and the fraction of variable sources is higher in their sample than in ours.
While we would have expected flux changes to be smeared out when 
considering average fluxes over long timescales, we still find 
11 variable sources, which suggests that flux variability is not uncommon 
at hard X-ray energies, although its amplitude is not 
dramatic. Therefore this should be taken as a warning when fitting together IBIS 
and BAT spectra.
In order to investigate possible changes also in spectral shape, we plot 
$\Gamma_{\rm IBIS}$ {\it vs.} $\Gamma_{\rm BAT}$ in Figure~\ref{gamma_bat_ibis}.
It is quite clear from this plot that there is overall good agreement between
the \emph{IBIS} and \emph{BAT} photon indices, suggesting that our sources
are not affected, on average, by changes in their spectral shapes. Only 
a few sources seem to deviate from the 1:1 line, suggesting a possible change 
in their photon index between the observations 
performed by the two instruments. However, the errors on both sides 
are large, thus indicating that a more appropriate analysis is needed. 
To highlight these objects, we applied the same procedure described 
above for the flux comparison: we find that at 3$\sigma$ confidence level 
only one source, namely 3C 273, is spectrally variable between the 
\emph{BAT} and the \emph{IBIS} measurements. This makes any combined \emph{BAT/IBIS}
analysis difficult to perform and therefore 3C 273
will not be discussed further.

To summarise the comparison between the two data sets, we can say
that there is, in general, a good agreement, 
apart from a few variable sources and a systematic difference in normalisation
between the two instruments which can be quantified, at this stage, at around 22\%.
As a final remark, we note that in the following sections we concentrate our efforts on the
\emph{BAT/IBIS} combined spectral analysis; for this reason, and
since we aim at a more in-depth study of our objects, 
sources for which no \emph{BAT} spectra are available will not be discussed further. 
Their simple power-law fits to the \emph{IBIS} data are nevertheless 
reported in Tables~\ref{ind_fit_type1} and \ref{ind_fit_type2}
for the sake of completeness.

\begin{table*}
\footnotesize
\centering
\caption{{\bf \emph{IBIS/BAT} individual fits - type 1 AGN}}
\label{ind_fit_type1}
\vspace{0.2cm}
\begin{tabular}{lcccrccrc}
\hline
{\bf Name}&{\bf N$_{\rm H}$}&{\bf $\bf \Gamma_{\bf \rm ISGRI}$}&{\bf F$_{\bf \rm 20-100keV}^{\bf \rm ISGRI}$}&{\bf $\bf \chi^2$ (d.o.f.)}&{\bf $\bf \Gamma_{\bf \rm BAT}$}&{\bf F$_{\bf \rm 20-100keV}^{\bf \rm BAT}$}&{\bf $\bf \chi^2$ (d.o.f.)}& {\bf Var.} \\
          &{\bf 10$^{\bf 22}$cm$^{\bf -2}$}& {\bf 10$^{\bf -11}$ (cgs)}&   &  &{\bf 10$^{\bf -11}$ (cgs)}   &      &  \\
\hline
IGR J00333+6122   &0.85&1.78$^{+0.31}_{-0.28}$&1.48$\pm$0.12& 3.8 (10) &2.17$^{+0.39}_{-0.35}$&0.93$\pm$0.11& 7.4 (5)&\\ 
QSO B0241+62      &0.32&1.89$^{+0.18}_{-0.17}$&4.88$\pm$0.26& 6.4 (10) &1.91$^{+0.07}_{-0.07}$&5.41$\pm$0.13& 7.9 (5)&\\
B3 0309+411       &0.13&2.07$^{+0.91}_{-0.77}$&1.90$\pm$0.65&13.8 (10) &1.38$^{+0.52}_{-0.54}$&1.06$\pm$0.18& 3.9 (5)&\\
3C 111            &0.43&1.82$^{+0.26}_{-0.25}$&10.48$\pm$0.90&10.7 (10)&1.97$^{+0.06}_{-0.06}$&7.25$\pm$0.14& 2.6 (5)&\\
LEDA 168563       &0.54&2.10$^{+0.47}_{-0.42}$&4.74$\pm$0.67& 9.3 (10) &1.92$^{+0.12}_{-0.12}$&3.79$\pm$0.15& 5.5 (5)&\\
4U 0517+17        &0.09&1.93$^{+0.18}_{-0.17}$&6.26$\pm$0.34& 8.2 (10) &2.05$^{+0.09}_{-0.09}$&5.58$\pm$0.17& 6.3 (5)&\\
MCG+08-11-011     &0.21&2.43$^{+0.64}_{-0.53}$&7.79$\pm$1.18& 5.5 (10) &2.02$^{+0.05}_{-0.05}$&8.19$\pm$0.16&29.8 (5)&\\
Mrk 6$^{\dagger}$ &4.79&1.97$^{+0.31}_{-0.28}$&4.09$\pm$0.36&16.1 (10) &1.89$^{+0.09}_{-0.09}$&3.79$\pm$0.13&10.9 (5)&\\
IGR J07597-3842   &0.60&2.26$^{+0.21}_{-0.20}$&3.47$\pm$0.21&16.4 (10) &2.03$^{+0.12}_{-0.11}$&3.18$\pm$0.13& 2.6 (5)&\\
ESO 209-12        &0.24&1.85$^{+0.27}_{-0.25}$&2.12$\pm$0.17& 6.2 (10) &2.13$^{+0.26}_{-0.23}$&1.40$\pm$0.13& 5.3 (5)&\\
FRL 1146          &0.32&2.27$^{+0.33}_{-0.30}$&1.98$\pm$0.18& 4.1 (10) &2.10$^{+0.21}_{-0.19}$&1.83$\pm$0.13& 11.1 (5)&\\
Swift J0917.2-6221&0.47&2.55$^{+0.48}_{-0.41}$&1.81$\pm$0.26&11.7 (10) &2.24$^{+0.19}_{-0.18}$&1.99$\pm$0.12& 4.3 (5)&\\
Swift J1038.8-4942&0.62&1.72$^{+0.35}_{-0.33}$&2.22$\pm$0.22& 4.6 (10) &1.75$^{+0.25}_{-0.24}$&1.64$\pm$0.16& 5.3 (5)&\\
NGC 3783          &1.15&2.23$^{+0.40}_{-0.36}$&12.69$\pm$0.16&19.8 (10)&1.96$^{+0.04}_{-0.04}$&11.70$\pm$0.12&20.1 (5)&\\
NGC 4151$^{\dagger}$&21.9&1.83$^{+0.04}_{-0.04}$&29.74$\pm$0.39&28.1 (10)&1.84$^{+0.01}_{-0.01}$&32.10$\pm$0.11&166.8 (5)&V, 0.81\\
Mrk 50            &0.12&2.76$^{+0.66}_{-0.57}$&1.21$\pm$0.20&16.0 (10)&1.92$^{+0.29}_{-0.27}$ &1.49$\pm$0.13& 2.7 (5)&\\
NGC 4593          &0.02&1.95$^{+0.10}_{-0.10}$&6.94$\pm$0.22&13.6 (10)&1.87$^{+0.08}_{-0.08}$ &5.37$\pm$0.16&10.6 (5)&\\
IGR J12415-5750   &0.30&1.99$^{+0.34}_{-0.31}$&1.74$\pm$0.18&12.7 (10)&1.73$^{+0.17}_{-0.16}$ &2.82$\pm$0.17& 1.7 (5)&V, 0.54\\
IGR J13109-5552   &0.46&1.67$^{+0.30}_{-0.28}$&2.37$\pm$0.19& 7.4 (10)&1.36$^{+0.29}_{-0.29}$ &1.83$\pm$0.20& 1.6 (5) &\\
MCG-06-30-015     &0.15&2.37$^{+0.27}_{-0.25}$&4.39$\pm$0.37&14.8 (10)&2.39$^{+0.12}_{-0.11}$ &4.12$\pm$0.16&17.6 (5)&\\
4U 1344-60$^{\dagger}$&42.6&2.08$^{+0.08}_{-0.08}$&7.35$\pm$0.18&26.8 (10)&2.01$^{+0.07}_{-0.07}$ &6.72$\pm$0.13&13.1 (5)&\\
IC 4329A          &0.35&1.98$^{+0.06}_{-0.06}$&20.95$\pm$0.42&43.7 (10)&2.04$^{+0.03}_{-0.03}$&18.20$\pm$0.18&110.9 (5)&\\
ESO 511-G030      &0.05&1.98$^{+0.42}_{-0.38}$&3.30$\pm$0.37&10.8 (10)&2.02$^{+0.20}_{-0.19}$ &2.78$\pm$0.19& 6.5 (5)&\\
IGR J16119-6036   &0.23&2.10$^{+0.26}_{-0.24}$&2.73$\pm$0.19& 6.9 (10)&2.15$^{+0.27}_{-0.25}$ &1.95$\pm$0.16& 5.4 (5)&\\
IGR J16185-5928   &0.24&1.75$^{+0.37}_{-0.33}$&1.89$\pm$0.19& 6.8 (10)&1.55$^{+0.41}_{-0.41}$ &1.35$\pm$0.07& 2.3 (5)&\\
IGR J16385-2057   &0.12&2.62$^{+0.70}_{-0.60}$&1.06$\pm$0.22&23.0 (10)& -& -& -&\\
IGR J16426+6536   & - &4.52$^{+2.13}_{-1.27}$&1.55$\pm$0.56&14.4 (10)& -& -& -&\\
IGR J16482-3036   &0.01&1.87$^{+0.16}_{-0.16}$&3.29$\pm$0.16& 5.6 (10)&1.95$^{+0.18}_{-0.18}$&3.27$\pm$0.03& 1.3 (5)&\\
IGR J16558-5203$^{\dagger}$&18.6&1.98$^{+0.18}_{-0.17}$&2.99$\pm$0.16&10.3 (10)&1.99$^{+0.19}_{-0.18}$&2.90$\pm$0.20& 4.2 (5)&\\
GRS 1734-292      &0.21&2.31$^{+0.04}_{-0.04}$&8.22$\pm$0.09&40.6 (10)&2.20$^{+0.08}_{-0.08}$&7.26$\pm$0.22& 8.7 (5)&\\
2E 1739.1-1210    &0.15&1.85$^{+0.21}_{-0.20}$&3.24$\pm$0.19& 8.2 (10)&2.08$^{+0.25}_{-0.23}$&2.57$\pm$0.23& 6.0 (5)&\\
IGR J17488-3253   &0.34&1.79$^{+0.06}_{-0.06}$&5.25$\pm$0.10&38.5 (10)&1.97$^{+0.25}_{-0.24}$&2.45$\pm$0.22& 4.1 (5)&V, 1.87\\
IGR J18027-1455   &0.30&1.80$^{+0.11}_{-0.11}$&4.99$\pm$0.15&27.5 (10)&1.85$^{+0.27}_{-0.26}$&2.33$\pm$0.23& 0.4 (5)&V, 1.87\\
IGR J18249-3243   &0.14&2.06$^{+0.57}_{-0.48}$&0.57$\pm$0.18&27.0 (10)&      -            & -&-&\\
IGR J18259-0706   &1.07&2.02$^{+0.36}_{-0.33}$&1.64$\pm$0.15&13.9 (10)&2.39$^{+0.49}_{-0.42}$&1.22$\pm$0.18& 1.8 (5)&\\
3C 390.3          &0.04&1.73$^{+0.25}_{-0.23}$&5.91$\pm$0.10& 7.9 (10)&2.01$^{+0.05}_{-0.05}$&6.42$\pm$0.13& 7.3 (5)&\\
2E 1853.7+1534    &0.39&2.24$^{+0.24}_{-0.22}$&2.51$\pm$0.41& 5.9 (10)&2.10$^{+0.24}_{-0.22}$&1.97$\pm$0.16& 2.7 (5)&\\
IGR J19378-0617   &0.15&1.93$^{+0.81}_{-0.68}$&1.82$\pm$0.31&19.8 (10)&2.33$^{+0.35}_{-0.31}$&1.54$\pm$0.15& 3.7 (5)&\\
NGC 6814          &0.13&1.91$^{+0.32}_{-0.29}$&5.35$\pm$0.48& 8.2 (10)&1.98$^{+0.11}_{-0.11}$&4.65$\pm$0.19& 5.5 (5)&\\
4C 74.26          &0.14&2.03$^{+0.79}_{-0.65}$&4.16$\pm$0.84&14.7 (10)&2.15$^{+0.11}_{-0.10}$&3.17$\pm$0.09&11.2 (5)&\\
S5 2116+81        &0.24&1.89$^{+0.74}_{-0.64}$&3.04$\pm$0.67&10.8 (10)&1.90$^{+0.16}_{-0.16}$&2.18$\pm$0.13& 5.6 (5)&\\
IGR J21247+5058$^{\dagger}$&7.76&1.98$^{+0.05}_{-0.05}$&12.51$\pm$0.17&26.9 (10)&2.00$^{+0.03}_{-0.03}$&12.30$\pm$0.12&24.0 (5)&\\
Swift J2127.4+5654&0.79&2.61$^{+0.22}_{-0.20}$&3.20$\pm$0.16&10.5 (10)&2.41$^{+0.15}_{-0.14}$&2.36$\pm$0.09& 3.6 (5)&\\
RX J2135.9+4728   &0.23&2.08$^{+0.40}_{-0.36}$&1.47$\pm$0.18&13.6 (10)&2.15$^{+0.27}_{-0.25}$&1.37$\pm$0.01& 2.5 (5)&\\
MR 2251-178       &2.14&1.79$^{+0.43}_{-0.38}$&6.31$\pm$0.86&10.5 (10)&2.01$^{+0.07}_{-0.07}$&6.24$\pm$0.12&15.2 (5)&\\
MCG-02-58-022     &0.04&1.74$^{+0.48}_{-0.43}$&4.52$\pm$0.69&12.2 (10)&1.95$^{+0.06}_{-0.06}$&6.98$\pm$0.14& 7.5 (5)&\\
\hline
\multicolumn{9}{l}{$^{\dagger}$: sources with complex absorption (partial covering), for which the highest value of
the column density is reported.}
\end{tabular}
\end{table*}

\begin{table*}
\footnotesize
\centering
\caption{{\bf \emph{IBIS/BAT} individual fits - type 2 AGN}}
\label{ind_fit_type2}
\vspace{0.2cm}
\begin{tabular}{lcccrccrc}
\hline
{\bf Name}&{\bf N$_{\rm \bf H}$}&{\bf $\bf \Gamma_{\rm \bf ISGRI}$}&{\bf F$_{\bf \rm 20-100keV}^{\bf \rm ISGRI}$}&{\bf $\bf \chi^2$ (d.o.f.)}&{\bf $\bf \Gamma_{\rm \bf BAT}$}&{\bf F$_{\bf \rm 20-100keV}^{\bf \rm BAT}$}&{\bf $\bf \chi^2$ (d.o.f.)}&{\bf Var.}\\
                &{\bf 10$^{\bf 22}$ cm$^{\bf -2}$}&     & {\bf 10$^{\bf -11}$ (cgs)}  &                 &       &{\bf 10$^{\bf -11}$ (cgs)}   & & \\
\hline
NGC 788           &30.2&1.88$^{+0.15}_{-0.16}$& 5.80$\pm$0.28& 5.7 (10)&1.84$^{+0.08}_{-0.08}$& 4.98$\pm$0.15&16.0 (5)&\\
NGC 1068          &1000&2.44$^{+0.78}_{-0.67}$& 2.20$\pm$0.29&11.9 (10)&3.09$^{+0.25}_{-0.24}$& 2.16$\pm$0.16& 4.3 (5)&\\
NGC 1142          &44.7&2.01$^{+0.23}_{-0.22}$& 6.21$\pm$0.42& 7.2 (10)&1.77$^{+0.08}_{-0.08}$& 5.60$\pm$0.16& 4.8 (5)&\\
NGC 1275          &0.12&3.61$^{+0.68}_{-0.55}$& 3.29$\pm$0.47&11.9 (10)&3.64$^{+0.11}_{-0.11}$& 3.79$\pm$0.07&10.6 (5)&\\
Mrk 3             & 100&1.72$^{+0.13}_{-0.13}$&10.90$\pm$0.45& 6.5 (10)&1.74$^{+0.05}_{-0.05}$& 7.90$\pm$0.15&12.3 (5)&\\
IGR J07565-4139   &0.59&1.89$^{+0.52}_{-0.47}$& 1.31$\pm$0.18& 8.3 (10)&     -            &  -  &  -    &   \\
MCG-05-23-16      &1.51&2.01$^{+0.16}_{-0.15}$&13.80$\pm$0.66&20.7 (10)&2.14$^{+0.03}_{-0.03}$&12.63$\pm$0.14&98.5 (5)&\\
IGR J09523-6231   &7.94&2.67$^{+0.57}_{-0.48}$& 1.35$\pm$0.16& 6.2 (10)&2.00$^{+0.39}_{-0.35}$& 0.96$\pm$0.13&  5.0(5)&\\
Swift J1009.3-4250&26.9&1.97$^{+0.39}_{-0.35}$& 2.87$\pm$0.30& 2.3 (10)&1.91$^{+0.21}_{-0.20}$& 2.04$\pm$0.14&11.2 (5)&\\
NGC 3281          & 200&1.84$^{+0.47}_{-0.41}$& 4.66$\pm$0.61&11.0 (10)&2.16$^{+0.08}_{-0.08}$& 5.42$\pm$0.16&23.4 (5)&\\
IGR J10404-4625   &2.67&1.98$^{+0.32}_{-0.30}$& 3.20$\pm$0.15& 8.1 (10)&2.06$^{+0.27}_{-0.25}$& 1.52$\pm$0.15& 3.3 (5)&V, 1.84\\
IGR J12026-5349   &3.31&2.02$^{+0.16}_{-0.15}$& 4.15$\pm$0.20& 7.7 (10)&2.12$^{+0.14}_{-0.13}$& 3.10$\pm$0.15& 9.9 (5)&\\
NGC 4388          &33.1&1.88$^{+0.05}_{-0.05}$&24.00$\pm$0.36&17.9 (10)&1.79$^{+0.02}_{-0.02}$&16.33$\pm$0.14&11.5 (5)&V, 1.29\\
NGC 4507          &28.8&1.90$^{+0.10}_{-0.10}$&15.60$\pm$0.48& 8.1 (10)&1.91$^{+0.04}_{-0.04}$&11.64$\pm$0.18&51.1 (5)&\\ 
LEDA 170194       &3.08&1.93$^{+0.34}_{-0.31}$& 4.06$\pm$0.38& 6.7 (10)&1.77$^{+0.20}_{-0.17}$& 2.61$\pm$0.18& 9.0 (5)&\\
NGC 4945          & 400&1.87$^{+0.03}_{-0.03}$&25.28$\pm$0.27&54.3 (10)&1.82$^{+0.03}_{-0.03}$&16.88$\pm$0.22&41.5 (5)&V, 1.31\\
IGR J13091+1137   &89&1.52$^{+0.34}_{-0.32}$& 3.90$\pm$0.46&19.7 (10)&1.75$^{+0.12}_{-0.10}$& 3.27$\pm$0.15&17.1 (5)&\\
Cen A             &37.0&1.86$^{+0.02}_{-0.02}$&59.00$\pm$0.29&10.2 (10)&1.90$^{+0.01}_{-0.01}$&83.73$\pm$0.18&51.0 (5)&V, 0.62\\
NGC 5252          &6.76&1.77$^{+0.48}_{-0.44}$& 5.45$\pm$0.77&15.2 (10)&1.86$^{+0.06}_{-0.06}$& 6.64$\pm$0.15& 3.0 (5)&\\
Circinus Galaxy  & 400&2.54$^{+0.03}_{-0.03}$&20.30$\pm$0.18&186.7 (10)&2.61$^{+0.03}_{-0.03}$&17.33$\pm$0.17&185.2 (5)&\\
NGC 5506          &3.40&2.50$^{+0.52}_{-0.44}$&15.80$\pm$0.22&12.1 (10)&2.07$^{+0.03}_{-0.03}$&15.31$\pm$0.17&198.5 (5)&\\
IGR J14515-5542   &0.50&1.86$^{+0.31}_{-0.28}$& 1.85$\pm$0.17&14.5 (10)&1.79$^{+0.18}_{-0.18}$& 2.79$\pm$0.18&10.8 (5)&\\
IC 4518A          &14.0&2.40$^{+0.27}_{-0.25}$& 2.06$\pm$0.18&15.2 (10)&2.46$^{+0.28}_{-0.25}$& 1.80$\pm$0.17&10.3 (5)&\\
IGR J16024-6107   &0.28&1.88$^{+0.53}_{-0.46}$    & 1.56$\pm$0.19      & 8.4 (10)  &      -           &  -   & -      & \\
IGR J16351-5806   & 370&2.12$^{+0.38}_{-0.34}$& 1.85$\pm$0.18&10.9 (10)&2.02$^{+0.40}_{-0.39}$& 1.70$\pm$0.20& 3.4 (5)&\\
NGC 6300          &23.0&2.17$^{+0.18}_{-0.17}$& 6.69$\pm$0.18&12.1 (10)&2.10$^{+0.07}_{-0.07}$& 6.11$\pm$0.17&30.3 (5)&\\
IGR J17513-2011   &0.71&1.84$^{+0.13}_{-0.13}$& 3.17$\pm$0.11&22.6 (10)&      -           &   -  &  -  &   \\
ESO 103-35        &20.0&1.36$^{+0.74}_{-0.52}$& 7.47$\pm$1.23& 7.6 (10)&2.16$^{+0.06}_{-0.06}$& 7.18$\pm$0.16&42.7 (5)&\\
Cyg A             &36.0&2.03$^{+0.08}_{-0.08}$& 8.49$\pm$0.20& 8.5 (10)&2.00$^{+0.05}_{-0.05}$& 8.43$\pm$0.13&25.0 (5)&\\
IGR J20186+4043   &10.0&2.24$^{+0.29}_{-0.27}$& 2.00$\pm$0.17&11.6 (10)&1.93$^{+0.20}_{-0.22}$& 1.80$\pm$0.15& 7.8 (5)&\\
NGC 7172          &8.50&2.01$^{+0.43}_{-0.21}$& 7.85$\pm$0.56& 5.1 (10)&1.85$^{+0.04}_{-0.04}$&10.47$\pm$0.16& 8.9 (5)&V, 0.66\\
IGR J23308+7120   &6.02&2.46$^{+0.67}_{-0.56}$& 1.01$\pm$0.16& 7.6 (10)&1.88$^{+0.65}_{-0.53}$& 0.71$\pm$0.12& 5.2 (5)&\\
IGR J23524+5842   & 6.30&1.58$^{+0.39}_{-0.35}$ & 1.13$\pm$0.13&14.9 (10)&            -      &  -  & -       &  \\
\hline
\end{tabular}
\end{table*}

\begin{table*}
\footnotesize
\centering
\caption{{\bf \emph{IBIS/BAT} individual fits - Blazars }}
\label{ind_fit_qso}
\vspace{0.2cm}
\begin{tabular}{lccrccrc}
\hline
{\bf Name}&{\bf $\bf \Gamma_{\bf \rm ISGRI}$}&{\bf F$_{\bf \rm 20-100keV}^{\bf \rm ISGRI}$}&{\bf $\bf \chi^2$ (d.o.f).}&{\bf $\Gamma_{\rm \bf BAT}$}&{\bf F$_{\bf \rm 20-100keV}^{\bf \rm BAT}$}&{\bf $\bf \chi^2$ (d.o.f.)}&{\bf Var.}\\
                &                                            & {\bf 10$^{\bf -11}$ (cgs)}          &                             &                            &{\bf 10$^{\bf -11}$ (cgs)}   &        \\
\hline
1ES 0033+595      &2.78$^{+0.25}_{-0.23}$& 1.94$\pm$0.10&15.1 (10)&2.80$^{+0.11}_{-0.05}$& 1.63$\pm$0.08& 6.8 (5)&\\  
QSO B0836+710     &1.65$^{+0.24}_{-0.22}$& 5.51$\pm$0.36&11.4 (10)&1.57$^{+0.09}_{-0.09}$& 4.03$\pm$0.12& 3.1 (5)&\\
4C 04.42          &1.00$^{+0.33}_{-0.32}$& 2.35$\pm$0.26& 6.8 (10)&1.31$^{+0.22}_{-0.22}$& 1.98$\pm$0.18& 5.9 (5)&\\
3C 273            &1.87$^{+0.04}_{-0.04}$&19.62$\pm$0.20&10.9 (10)&1.72$^{+0.02}_{-0.02}$&25.30$\pm$0.15&15.0 (5)&V, 0. 69\\
3C 279            &1.54$^{+0.32}_{-0.30}$& 2.40$\pm$0.24&10.1 (10)&1.49$^{+0.26}_{-0.25}$& 1.91$\pm$0.17& 1.2 (5)&\\
Swift J1656.3-3302&1.69$^{+0.17}_{-0.18}$& 2.67$\pm$0.13& 3.2 (10)&1.59$^{+0.16}_{-0.14}$& 3.99$\pm$0.24& 7.2 (5)&V, 0.59\\ 
PKS 1830-211      &1.66$^{+0.14}_{-0.14}$& 3.99$\pm$0.16&14.0 (10)&1.49$^{+0.13}_{-0.13}$& 4.90$\pm$0.24& 3.9 (5)&\\  
BL Lac            &1.81$^{+0.37}_{-0.34}$& 2.48$\pm$0.25& 4.7 (10)&1.87$^{+0.23}_{-0.22}$& 1.69$\pm$0.13& 4.7 (5)&\\ 
\hline 
\end{tabular}
\end{table*}

\section{\emph{BAT} and \emph{IBIS} combined spectral analysis: 
cross-calibration and photon index distributions}\label{cross_calib}

After the initial individual fits of \emph{IBIS} and \emph{BAT} spectra,
we combined the two datasets and fit them together in \texttt{XSPEC}; 
this is justified by the lack of significant spectral variability,
as discussed in the previous section. To take into account those few cases
affected by flux changes and the likely mismatch between the \emph{IBIS} and
\emph{BAT} spectra discussed above, we introduced a 
cross-calibration constant, \emph{C}, in the fit.
{\bf The model employed is a simple power-law, absorbed by an intrinsic column
density, as done for the individual fits.}
The results of this analysis are reported in Tables~\ref{sim_fit_type1}, 
\ref{sim_fit_type2} and \ref{sim_fit_qso}, where again we list source name,
combined photon index $\Gamma$, cross-calibration constant \emph{C}, \emph{IBIS}
and \emph{BAT} 20--100\,keV flux, $\chi^2$ and degrees of freedom.

\begin{figure}
\centering
\includegraphics[width=1.0\linewidth]{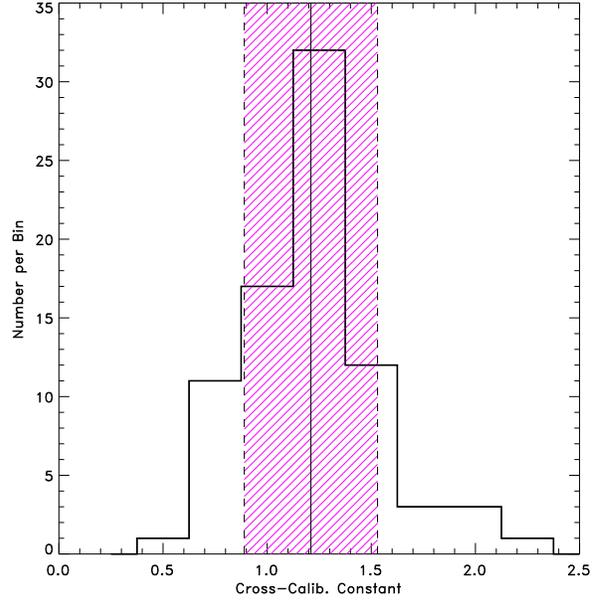}
\caption{Histogram of the cross-calibration constant between \emph{IBIS} and \emph{BAT}
for the sample of 80 AGN with combined data. The vertical solid line represents the average 
value of \emph{C}, while the dashed lines and the hatched area represent the parameter dispersion.}
\label{c_bat_ibis}
\end{figure}

In Figure~\ref{c_bat_ibis} the distribution of the cross-calibration
constants is shown.  The average value of \emph{C} is 1.22$\pm$0.03
(solid vertical line in Figure~\ref{c_bat_ibis}),
with a standard deviation of 0.32 (dashed lines and hatched area).
 
\begin{figure}
\centering
\includegraphics[width=1.05\linewidth]{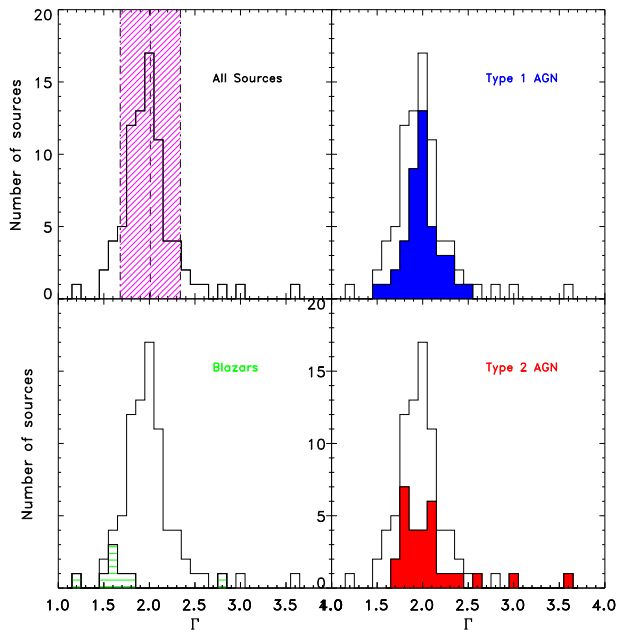}
\caption{\emph{Top-left Panel:} Photon index distribution for the total sample;
the shaded area represents the mean and standard deviation boundaries. \emph{Top-right
Panel}: Photon index distribution of type 1 sources.
\emph{Bottom-left Panel}: Photon index distribution of Blazars. 
\emph{Bottom-right Panel}: Photon index distribution of type 2 sources.
In all panels the black empty histogram represents the entire sample of 80 AGN while the coloured areas represent the specific
class of objects studied.}
\label{g_bat_ibis_class}
\end{figure}

Figure~\ref{g_bat_ibis_class} shows the photon
index distribution for all 80 AGN in the sample and for each class of objects
separately. Considering the entire sample, we find an average photon index of 
2.01$\pm$0.04, with a standard deviation of 0.33. For type 1 
and type 2 AGN we find $<$$\Gamma$$>$=2.00$\pm$0.03 ($\sigma$=0.19) and
$<$$\Gamma$$>$=2.10$\pm$0.08 ($\sigma$=0.41) respectively, while for Blazars the average
photon index is slightly flatter ($<$$\Gamma$$>$=1.73$\pm$0.19, $\sigma$=0.50).
Our results show that, at high energies, the primary continuum has a 
{\textquotedblleft{canonical}\textquotedblright} shape, at least in non-Blazars sources, and further indicate 
that type 1 and type 2 AGN have overlapping ranges of photon indices, thus suggesting
that the high energy emission is produced by the same mechanism. 

\citet{beckmann09}, who analysed a sample of 
148 objects employing \emph{INTEGRAL/IBIS} data, find that the
underlying continuum of type 1 and 2 AGN is generally steep as found here but not 
exactly overlapping. In fact, they report
$<$$\Gamma$$>$=1.97$\pm$0.03 for Seyfert 1-1.5 and $<$$\Gamma$$>$=1.88$\pm$0.02 for Seyfert 2.
Similarly, \citet{burlon11}, using \emph{Swift/BAT} data of 199 AGN, obtained 
$<$$\Gamma$$>$=2.07$\pm$0.03 ($\sigma$=0.27) for unobscured sources
and $<$$\Gamma$$>$=1.92$\pm$0.02 ($\sigma$=0.25) for obscured objects;
contrary to our results, both these studies suggest that type 2/obscured
sources could have a flatter spectral index than type 1/unobscured objects.
The difference in the hard X-ray spectral slope between type 1 and 2 
has been a point of discussion in the literature for some time now;
it was first noted by \citet{zdziarski95} using \emph{Ginga} and \emph{CGRO/OSSE} data and later confirmed by other
studies \citep{deluit03, beckmann06}.  This could be due to the fact that
absorption may play a role even above 10\,keV if the column density is very high, i.e. close to or
in the Compton regime. The resulting effect is a flattening of the  
hard X-ray spectra of type 2 AGN; should absorption be properly accounted for, 
no major difference should be evident in the two samples. Indeed, if absorption is not
considered in either class, the spectra of Seyfert 2 are flatter, although still consistent with
those of Seyfert 1, leaving the issue still open.
   
We can also compare the mean photon index obtained for our type 1 AGN 
with that reported in the 2--10\,keV band for the CAIXA catalogue \citep{Bianchi:2009}.
This catalogue consists of all radio-quiet, X-ray unobscured 
objects observed by \emph{XMM-Newton} in targeted observations, and lists 156 sources.
Albeit limited to only type 1 objects, this soft X-ray catalogue is the most compatible with our sample as it 
contains bright AGN similar to ours. The average 2--10\,keV photon index for the CAIXA sample is  
$<$$\Gamma$$>$=1.73 ($\sigma$=0.45), a much flatter value than 
that found by our analysis. Even when the CAIXA AGN are fitted 
with a model that takes into account Compton reflection, the average
photon index does not steepen enough to meet our value, being 1.78 ($\sigma$=0.46);
this strongly suggests that an extra component, such as a cut-off at high energies, is required
to model AGN spectra. We note that other soft X-ray selected samples of AGN 
report a steeper average photon index ($\Gamma$$\sim$2; see, for example, \citealt{corral11}),
but are generally made of weaker objects where spectral analysis is more difficult and uncertain. 

\begin{table*}
\footnotesize
\centering
\caption{{\bf \emph{BAT/IBIS} simultaneous fits - type 1 AGN }}
\label{sim_fit_type1}
\vspace{0.2cm}
\begin{tabular}{lccccr}
\hline
{\bf Name}&{\bf $\bf \Gamma$}&{\bf C$_{\rm \bf IBIS/BAT}$}&{\bf F$_{\bf \rm 20-100keV}^{\bf \rm IBIS}$}&{\bf F$_{\bf \rm 20-100keV}^{\bf \rm BAT}$}&{\bf $\bf \chi^2$ (d.o.f.)}\\
                  &                           &          &   {\bf 10$^{\bf -11}$ (cgs)}&{\bf 10$^{\bf -11}$ (cgs)}&\\
\hline
IGR J00333+6122   &1.95$^{+0.24}_{-0.23}$&1.41$^{+0.38}_{-0.28}$& 1.42$\pm$0.11& 1.02$\pm$0.12&13.2 (16)\\
QSO B0241+62      &1.91$^{+0.06}_{-0.06}$&0.91$^{+0.08}_{-0.08}$& 4.85$\pm$0.27& 5.42$\pm$0.13&14.4 (16)\\
B3 0309+411B      &1.59$^{+0.44}_{-0.43}$&2.05$^{+1.22}_{-0.90}$& 2.04$\pm$0.69& 1.02$\pm$0.17&19.1 (16)\\
3C 111            &1.96$^{+0.06}_{-0.06}$&1.42$^{+0.19}_{-0.18}$&10.10$\pm$0.91& 7.27$\pm$0.14&14.0 (16)\\
LEDA 168563       &1.94$^{+0.12}_{-0.11}$&1.33$^{+0.30}_{-0.29}$& 4.92$\pm$0.74& 3.77$\pm$0.15&15.2 (16)\\
4U 0517+17        &2.02$^{+0.08}_{-0.08}$&1.11$^{+0.11}_{-0.10}$& 6.13$\pm$0.31& 5.63$\pm$0.17&15.4 (16)\\
MCG+08-11-011     &2.02$^{+0.05}_{-0.05}$&1.08$^{+0.22}_{-0.22}$& 8.67$\pm$1.13& 8.17$\pm$0.16&36.9 (16)\\
Mrk 6$^{\dagger}$  &1.89$^{+0.09}_{-0.09}$&1.12$^{+0.15}_{-0.14}$& 4.18$\pm$0.33& 3.78$\pm$0.13&27.2 (16)\\
IGR J07597-3842   &2.09$^{+0.10}_{-0.10}$&1.20$^{+0.13}_{-0.12}$& 3.63$\pm$0.22& 3.10$\pm$0.12&21.7 (16)\\
ESO 209-12        &2.01$^{+0.18}_{-0.17}$&1.41$^{+0.27}_{-0.23}$& 2.04$\pm$0.16& 1.48$\pm$0.13&13.2 (16)\\ 
FRL 1146          &2.15$^{+0.17}_{-0.16}$&1.17$^{+0.22}_{-0.19}$& 2.04$\pm$0.18& 1.78$\pm$0.12&15.8 (16)\\
Swift J0917.2-6221&2.29$^{+0.18}_{-0.16}$&1.03$^{+0.21}_{-0.19}$& 1.95$\pm$0.25& 1.94$\pm$0.12&17.2 (16)\\
Swift J1038.8-4942&1.74$^{+0.20}_{-0.19}$&1.37$^{+0.32}_{-0.26}$& 2.21$\pm$0.22& 1.64$\pm$0.16& 9.9 (16)\\
NGC 3783          &1.96$^{+0.04}_{-0.04}$&1.18$^{+0.20}_{-0.20}$& 1.35$\pm$0.16& 1.16$\pm$0.02&41.4 (16)\\
NGC 4151$^{\dagger}$&1.84$^{+0.01}_{-0.01}$&0.94$^{+0.02}_{-0.02}$&29.70$\pm$0.36&32.10$\pm$0.11&195.1 (16)\\
Mrk 50            &2.11$^{+0.28}_{-0.26}$&1.09$^{+0.33}_{-0.28}$& 1.46$\pm$0.22& 1.37$\pm$0.12&23.4 (16)\\
NGC 4593          &1.90$^{+0.06}_{-0.06}$&1.35$^{+0.09}_{-0.08}$& 7.02$\pm$0.21& 5.31$\pm$0.16&25.2 (16)\\
IGR J12415-5750   &1.78$^{+0.15}_{-0.14}$&0.67$^{+0.12}_{-0.11}$& 1.82$\pm$0.20& 2.77$\pm$0.17&15.9 (16)\\
IGR J13109-5552   &1.52$^{+0.20}_{-0.20}$&1.37$^{+0.32}_{-0.25}$& 2.39$\pm$0.19& 1.78$\pm$0.19&10.6 (16)\\
MCG-06-30-015     &2.39$^{+0.11}_{-0.10}$&1.08$^{+0.13}_{-0.13}$& 4.37$\pm$0.30& 4.12$\pm$0.16&32.5 (16)\\
4U 1344-60$^{\dagger}$&2.04$^{+0.05}_{-0.05}$&1.14$^{+0.06}_{-0.05}$& 7.45$\pm$0.15& 6.63$\pm$0.13&41.1 (16)\\
IC 4329A          &2.03$^{+0.02}_{-0.02}$&1.15$^{+0.04}_{-0.04}$&20.70$\pm$0.41&18.30$\pm$0.18&156.7 (16)\\
ESO 511-G030      &2.01$^{+0.18}_{-0.17}$&1.20$^{+0.25}_{-0.23}$& 3.27$\pm$0.36& 2.79$\pm$0.19&17.3 (16)\\
IGR J16119-6036   &2.13$^{+0.18}_{-0.17}$&1.40$^{+0.24}_{-0.20}$& 2.71$\pm$0.19& 1.97$\pm$0.16&12.4 (16)\\
IGR J16185-5298   &1.67$^{+0.27}_{-0.25}$&1.47$^{+0.46}_{-0.33}$& 1.90$\pm$0.19& 1.31$\pm$0.20& 9.5 (16)\\
IGR J16482-3036   &1.91$^{+0.12}_{-0.12}$&1.01$^{+0.12}_{-0.11}$& 3.27$\pm$0.16& 3.30$\pm$0.03& 7.1 (16)\\
IGR J16558-5203$^{\dagger}$&1.98$^{+0.13}_{-0.12}$&1.05$^{+0.13}_{-0.12}$& 2.99$\pm$0.15& 2.90$\pm$0.20&14.5 (16)\\
GRS 1734-292      &2.29$^{+0.03}_{-0.03}$&1.20$^{+0.05}_{-0.05}$& 8.27$\pm$0.08& 7.07$\pm$0.21&53.2 (16)\\
2E 1739.1-1210    &1.95$^{+0.16}_{-0.15}$&1.22$^{+0.19}_{-0.16}$& 3.17$\pm$0.19& 2.66$\pm$0.24&15.6 (16)\\
IGR J17488-3253   &1.81$^{+0.06}_{-0.06}$&2.11$^{+0.28}_{-0.23}$& 5.24$\pm$0.10& 2.52$\pm$0.23&44.0 (16)\\
IGR J18027-1455   &1.81$^{+0.10}_{-0.10}$&2.14$^{+0.34}_{-0.26}$& 4.98$\pm$0.15& 2.36$\pm$0.24&28.0 (16)\\
IGR J18259-0706   &2.17$^{+0.29}_{-0.27}$&1.22$^{+0.36}_{-0.26}$& 1.58$\pm$0.14& 1.32$\pm$0.20&16.9 (16)\\
3C 390.3          &1.99$^{+0.05}_{-0.05}$&0.87$^{+0.10}_{-0.10}$& 5.48$\pm$0.38& 6.45$\pm$0.13&18.4 (16)\\
2E 1853.7+1534    &2.17$^{+0.17}_{-0.16}$&1.36$^{+0.23}_{-0.19}$& 2.55$\pm$0.16& 1.91$\pm$0.15& 9.2 (16)\\
IGR J19378-0617   &2.29$^{+0.30}_{-0.28}$&1.14$^{+0.38}_{-0.32}$& 1.74$\pm$0.29& 1.56$\pm$0.16&22.5 (16)\\
NGC 6814          &1.97$^{+0.11}_{-0.10}$&1.15$^{+0.18}_{-0.17}$& 5.28$\pm$0.47& 4.66$\pm$0.19&13.8 (16)\\
4C 74.26          &2.16$^{+0.11}_{-0.10}$&1.36$^{+0.36}_{-0.35}$& 4.21$\pm$0.76& 3.17$\pm$0.09&25.0 (16)\\
S5 2116+81        &1.90$^{+0.16}_{-0.15}$&1.42$^{+0.46}_{-0.43}$& 3.03$\pm$0.61& 2.18$\pm$0.13&16.4 (16)\\
IGR J21247+5058$^{\dagger}$&1.99$^{+0.02}_{-0.02}$&1.03$^{+0.03}_{-0.03}$&12.50$\pm$0.12&12.30$\pm$0.12&51.4 (16)\\
Swift J2127.4+5654&2.48$^{+0.12}_{-0.12}$&1.50$^{+0.16}_{-0.15}$& 3.33$\pm$0.18& 2.27$\pm$0.09&15.7 (16)\\
RX J2135.9+4728   &2.12$^{+0.22}_{-0.21}$&1.07$^{+0.25}_{-0.21}$& 1.45$\pm$0.16& 1.38$\pm$0.14&16.1 (16)\\
MR 2251-178       &2.00$^{+0.07}_{-0.06}$&0.97$^{+0.21}_{-0.21}$& 5.93$\pm$0.89& 6.25$\pm$0.12&26.3 (16)\\
MCG-02-58-022     &1.95$^{+0.06}_{-0.06}$&0.62$^{+0.14}_{-0.13}$& 4.25$\pm$0.06& 6.99$\pm$0.14&20.2 (16)\\
\hline
\multicolumn{6}{l}{$^{\dagger}$: see table \ref{ind_fit_type1}.}
\end{tabular}
\end{table*}

\begin{table*}
\footnotesize
\centering
\caption{{\bf \emph{BAT/IBIS} simultaneous fits - type 2 AGN }}
\label{sim_fit_type2}
\vspace{0.2cm}
\begin{tabular}{lccccr}
\hline
{\bf Name} & {\bf $\bf \Gamma$} & {\bf C$_{\rm \bf IBIS/BAT}$} & {\bf F$^{\rm \bf IBIS}_{\rm \bf 20-100keV}$}&{\bf F$^{\rm \bf BAT}_{\rm \bf 20-100 keV}$}& {\bf $\bf \chi^2$ (d.o.f.)}\\
  &  &  & {\bf $10^{\bf -11}$ erg cm$^{\bf -2}$ s$^{\bf -1}$} &  {\bf $10^{\bf -11}$ erg cm$^{\bf -2}$ s$^{\bf -1}$} &  \\
\hline
NGC 788           &1.85$^{+0.07}_{-0.07}$&1.19$^{+0.10}_{-0.10}$& 5.81$\pm$0.29& 4.97$\pm$0.15& 21.7 (16)\\
NGC 1068          &3.04$^{+0.23}_{-0.22}$&0.96$^{+0.22}_{-0.20}$& 2.05$\pm$0.33& 2.19$\pm$0.17& 15.4 (16)\\
NGC 1142          &1.80$^{+0.07}_{-0.07}$&1.19$^{+0.12}_{-0.12}$& 6.50$\pm$0.39& 5.56$\pm$0.16& 14.9 (16)\\
NGC 1275          &3.64$^{+0.11}_{-0.10}$&0.82$^{+0.13}_{-0.13}$& 3.02$\pm$0.33& 3.79$\pm$0.07& 22.5 (16)\\ 
Mrk 3             &1.74$^{+0.05}_{-0.05}$&1.20$^{+0.08}_{-0.08}$& 9.35$\pm$1.04& 7.91$\pm$0.15& 18.9 (16)\\
MCG-05-23-16      &2.14$^{+0.03}_{-0.03}$&1.07$^{+0.08}_{-0.08}$&13.30$\pm$0.66&12.66$\pm$0.14&121.0 (16)\\
IGR J09523-6231   &2.28$^{+0.33}_{-0.29}$&1.84$^{+0.63}_{-0.45}$& 1.51$\pm$0.18& 0.84$\pm$0.11& 14.5 (16)\\
Swift J1009.3-4250&1.92$^{+0.18}_{-0.17}$&1.45$^{+0.28}_{-0.25}$& 2.90$\pm$0.29& 2.03$\pm$0.14& 13.5 (16)\\
NGC 3281          &2.15$^{+0.08}_{-0.08}$&0.80$^{+0.17}_{-0.17}$& 4.28$\pm$0.60& 5.44$\pm$0.16& 35.6 (16)\\
IGR J10404-4625   &2.03$^{+0.20}_{-0.19}$&2.10$^{+0.46}_{-0.38}$& 3.16$\pm$0.32& 1.54$\pm$0.15& 11.2 (16)\\
IGR J12026-5349   &2.08$^{+0.10}_{-0.10}$&1.32$^{+0.14}_{-0.12}$& 4.08$\pm$0.16& 3.15$\pm$0.16& 18.2 (16)\\
NGC 4388          &1.80$^{+0.02}_{-0.02}$&1.53$^{+0.04}_{-0.04}$&24.42$\pm$0.34&16.23$\pm$0.14& 36.2 (16)\\       
NGC 4507          &1.91$^{+0.04}_{-0.04}$&1.36$^{+0.07}_{-0.07}$&15.58$\pm$0.47&11.65$\pm$0.18& 59.3 (16)\\       
LEDA 170194       &1.82$^{+0.16}_{-0.15}$&1.66$^{+0.29}_{-0.26}$& 4.17$\pm$0.33& 2.56$\pm$0.18& 16.1 (16)\\
NGC 4945          &1.84$^{+0.02}_{-0.02}$&1.56$^{+0.04}_{-0.03}$&25.97$\pm$0.26&16.81$\pm$0.22& 99.0 (16)\\
IGR J13091+1137   &1.74$^{+0.11}_{-0.11}$&1.15$^{+0.22}_{-0.21}$& 3.73$\pm$0.45& 3.29$\pm$0.13& 38.0 (16)\\
Cen A             &1.89$^{+0.01}_{-0.01}$&0.71$^{+0.01}_{-0.01}$&58.48$\pm$0.03&83.86$\pm$0.18& 75.5 (16)\\
NGC 5252          &1.85$^{+0.06}_{-0.06}$&0.82$^{+0.17}_{-0.17}$& 5.36$\pm$0.75& 6.64$\pm$0.13& 18.2 (16)\\
Circinus Galaxy   &2.58$^{+0.02}_{-0.02}$&1.18$^{+0.02}_{-0.02}$&20.14$\pm$0.18&17.48$\pm$0.17&378.9 (16)\\
NGC 5506          &2.07$^{+0.03}_{-0.03}$&1.17$^{+0.22}_{-0.22}$&17.62$\pm$2.11&15.30$\pm$0.15&213.1 (16)\\
IGR J14515-5542   &1.81$^{+0.16}_{-0.15}$&0.68$^{+0.12}_{-0.11}$& 1.87$\pm$0.19& 2.78$\pm$0.18& 25.3 (16)\\
IC 4518A          &2.43$^{+0.19}_{-0.18}$&1.15$^{+0.21}_{-0.17}$& 2.05$\pm$0.16& 1.83$\pm$0.16& 25.5 (16)\\
IGR J16351-5806   &2.05$^{+0.28}_{-0.26}$&1.13$^{+0.28}_{-0.22}$& 1.88$\pm$0.19& 1.69$\pm$0.18& 13.7 (16)\\
NGC 6300          &2.11$^{+0.07}_{-0.07}$&1.14$^{+0.10}_{-0.10}$& 6.78$\pm$0.34& 6.09$\pm$0.18& 42.8 (16)\\
ESO 103-35        &2.13$^{+0.06}_{-0.06}$&0.88$^{+0.24}_{-0.23}$& 6.21$\pm$0.99& 7.19$\pm$0.14& 58.0 (16)\\
Cyg A             &2.01$^{+0.04}_{-0.04}$&1.03$^{+0.05}_{-0.04}$& 8.54$\pm$0.17& 8.40$\pm$0.13& 33.8 (16)\\
IGR J20186+4043   &2.05$^{+0.17}_{-0.16}$&1.23$^{+0.23}_{-0.19}$& 2.08$\pm$0.17& 1.73$\pm$0.14& 21.8 (16)\\
NGC 7172          &1.86$^{+0.04}_{-0.04}$&0.80$^{+0.08}_{-0.08}$& 8.21$\pm$0.49&10.45$\pm$0.16& 15.6 (16)\\
IGR J23308+7120   &2.19$^{+0.48}_{-0.40}$&1.83$^{+0.80}_{-0.57}$& 1.09$\pm$0.17& 0.61$\pm$0.10& 14.2 (16)\\
\hline 
\end{tabular}
\end{table*}

\begin{table*}
\footnotesize
\centering
\caption{{\bf \emph{BAT/IBIS} simultaneous fits - Blazars }}
\label{sim_fit_qso}
\vspace{0.2cm}
\begin{tabular}{lccccr}
\hline
{\bf Name} & {\bf $\bf \Gamma$} & {\bf C$_{\rm \bf IBIS/BAT}$} & {\bf F$^{\rm \bf IBIS}_{\rm \bf 20-100keV}$}&{\bf F$^{\rm \bf BAT}_{\rm \bf 20-100 keV}$}& {\bf $\bf \chi^2$ (d.o.f.)}\\
  &  &  & {\bf $10^{\bf -11}$ erg cm$^{\bf -2}$ s$^{\bf -1}$} &  {\bf $10^{\bf -11}$ erg cm$^{\bf -2}$ s$^{\bf -1}$} &  \\
\hline
1ES 0033+595      &2.79$^{+0.16}_{-0.15}$&1.20$^{+0.14}_{-0.13}$&1.93$\pm$0.10&1.64$\pm$0.08&21.9 (16)\\
QSO B0836+710     &1.58$^{+0.08}_{-0.08}$&1.41$^{+0.17}_{-0.16}$&5.58$\pm$0.37&4.02$\pm$0.12&14.8 (16)\\
4C 04.42          &1.20$^{+0.20}_{-0.17}$&1.17$^{+0.27}_{-0.23}$&2.32$\pm$0.25&2.01$\pm$0.18&14.3 (16)\\ 
3C 279            &1.52$^{+0.20}_{-0.19}$&1.28$^{+0.29}_{-0.24}$&2.41$\pm$0.24&1.90$\pm$0.17&11.4 (16)\\ 
Swift J1656.3-3302&1.64$^{+0.11}_{-0.11}$&0.69$^{+0.08}_{-0.07}$&2.68$\pm$0.13&3.96$\pm$0.24&10.8 (16)\\   
PKS 1830-211      &1.58$^{+0.09}_{-0.08}$&0.85$^{+0.08}_{-0.07}$&4.02$\pm$0.16&4.82$\pm$0.24&20.4 (16)\\  
BL Lac            &1.83$^{+0.21}_{-0.17}$&1.47$^{+0.32}_{-0.27}$&2.47$\pm$0.25&1.71$\pm$0.14& 9.4 (16)\\
\hline 
\end{tabular}
\end{table*}

\subsection{Spectral Complexity}\label{spec_compl}

Although a simple power-law model describes sufficiently well the 
data points for the majority of our sources, in some cases
it does not fit the data properly. In fact, as can be seen from the fits reported 
in Tables~\ref{sim_fit_type1} and \ref{sim_fit_type2}, 28 sources or $\sim$30\% of the sample (14 type
1 and 14 type 2 objects) have reduced $\chi^2$$\geq$1.5, suggesting 
that a more complex model might be needed to describe their spectra. 
We have adopted a $\chi^2$ threshold value of 
1.5 to discriminate between well and bad fitted objects, as there is only a probability of 10\%
of exceeding this value by chance, i.e. no more than 8 sources in the entire sample are expected to have a bad
fit. 

In the case of these 28 AGN, we have re-fitted the combined \emph{BAT/IBIS} data using the \texttt{pexrav}
model in \texttt{XSPEC}; this model takes into account
features such as the high energy cut-off (E$_{\rm cut}$) and/or the reflection fraction (R), 
which can both be present in the energy range probed by the present study. The reflection 
component depends on the inclination angle {\it i} between the axis perpendicular to 
the accretion disk and the line of sight: the smaller the inclination angle, the larger the resulting
reflection component. Since the quality of the data does not allow to constrain both 
reflection and inclination at the same time, we have fixed cos({\it i}) = 0.45 for both type 1 and 2 AGN.

In the first instance, we left all relevant parameters free to vary and report in the 
first row of Tables~\ref{pex_fit1}/\ref{pex_fit1_cont} and \ref{pex_fit2}/\ref{pex_fit2_cont} 
the results of these fits; for each source, 
the tables list photon index, high energy cut-off, reflection fraction, cross-calibration constant,
$\chi^2$ and degrees of freedom, as well as the probability of a fit improvement
with respect to the simple power-law model reported in Tables~\ref{sim_fit_type1} and \ref{sim_fit_type2}. 
Although the fit improvement is significant in most cases, it is evident that it is quite difficult 
to put firm constraints on all free parameters, i.e. power-law photon index, high 
energy cut-off and reflection, at the same time. This is mainly due to the lack of 
data below 17\,keV and the fact that unfortunately the parameters of interest 
are closely linked in the fitting procedure. 
Out of 28 AGN analysed, only in two cases (NGC 4151 and IC 4329A)
it is possible to put constraints on all 3 parameters at the same time. In most other cases,  
only two out of three parameters analysed could be constrained, specifically the 
photon index with either the reflection 
(in the minority of the sources) or the high energy cut-off (in the majority of the objects).
There are nevertheless some AGN in which neither R or E$_{\rm cut}$ could be constrained at all. 
This suggests that if one wants to get some spectral information from hard X-ray spectra alone,  
some assumptions must be made on at least one of the 3 parameters involved in the spectral analysis.
For example, among type 1 sources, IGR J21247+5058 is a well studied source
which is known to have little reflection \citep{Molina:2007, tazaki10}; in this case, if we fix the
reflection fraction to zero and leave both the photon index and the cut-off
energy as free parameters, both quantities are indeed well constrained 
(see Table~\ref{pex_fit1}, \ref{pex_fit1_cont}) and are compatible with previous studies. On the other hand, the
Circinus Galaxy (a Compton thick AGN) has a well-known cut-off energy located at around 50\,keV,
measured by several instruments such as \emph{BeppoSAX} and \emph{XTE} (e.g. \citealt{yang09}): 
fixing E$_{\rm cut}$ at this value, we are able to constrain both $\Gamma$ to be 1.80$^{+0.03}_{-0.03}$ 
and R to be 0.59$^{+0.29}_{-0.24}$. Unfortunately not all our sources are well studied at high energies
as these two objects and so a more general approach has been adopted in order to deal with all 28 AGN.
We therefore choose to fix the reflection component to be 0, 1, or 2; the higher value of R
allows for the extra reflection often observed in AGN and likely due to a more complex 
or peculiar geometry, such as one in which a molecular torus is present alongside an accretion disk,
more radiation is emitted toward the reflector than toward the observer or light 
bending effects are at work \citep{Molina:2009}.

The results obtained with fixed values of reflection are  reported in Tables~\ref{pex_fit1}/\ref{pex_fit1_cont} 
and \ref{pex_fit2}/\ref{pex_fit2_cont},
where again a significant fit improvement (close to 99\% or more in at least 
one of the three cases of R analysed) is evident in many objects, when comparing
these fits with single power-law ones; the improvement occurs in 75\% of the sources,
therefore leaving only a handful of AGN where the improvement is not as significant. 
In a couple of AGN, we could not obtain any constraints on the cut-off energy,
presumably because it is located at higher energies than those probed by \emph{IBIS/BAT}. 
Finally, in a small number of objects, i.e. 10 out of 28, the reduced $\chi^2$ obtained using 
the \texttt{pexrav} model is still above the threshold of 1.5 adopted in the present study, 
meaning that in these sources we have not yet achieved a good fit. 
These 10 AGN (4 type 1 and 5 type 2) are highlighted in 
Tables~\ref{pex_fit1}, \ref{pex_fit1_cont} and \ref{pex_fit2}, \ref{pex_fit2_cont} 
and are discussed in more detail in the next section.

\subsection{Further spectral considerations}

\begin{figure}
\centering
\includegraphics[scale=0.3,angle=-90]{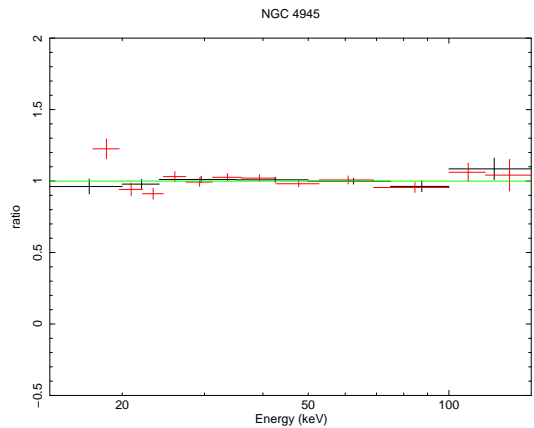}
\caption{Best fit model-to-\emph{IBIS/BAT} data (\texttt{pexrav}, R free) ratio for NGC 4945 (see Table~\ref{pex_fit2}). 
\emph{IBIS} data are in red while \emph{BAT} data are in black. As can be seen from the plot,
the first channel of the \emph{IBIS} spectrum is causing the fit to have a $\Delta$$\chi^2$$>$1.5.}
\label{ngc4945}
\end{figure}

\begin{figure}
\centering
\includegraphics[scale=0.3,angle=-90]{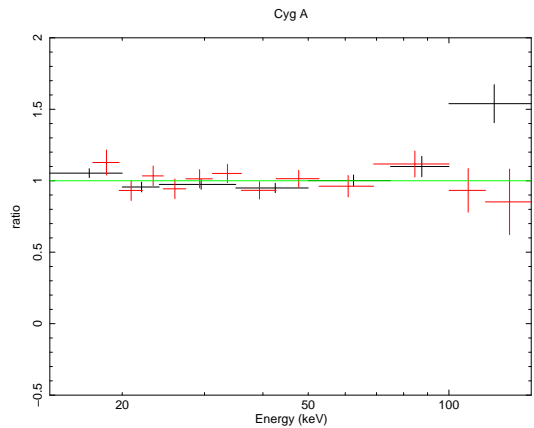}
\caption{Best fit model-to-\emph{IBIS/BAT} data ratio (\texttt{pexrav}, R=0) for Cyg A (see Table~\ref{pex_fit2}). 
\emph{IBIS} data are in red while \emph{BAT} data are in black.
As can be seen from the plot, the last channel of the \emph{BAT} spectrum is causing the fit to have a $\Delta$$\chi^2$$>$1.5.}
\label{cyga}
\end{figure}

\begin{figure}
\centering
\includegraphics[scale=0.3,angle=-90]{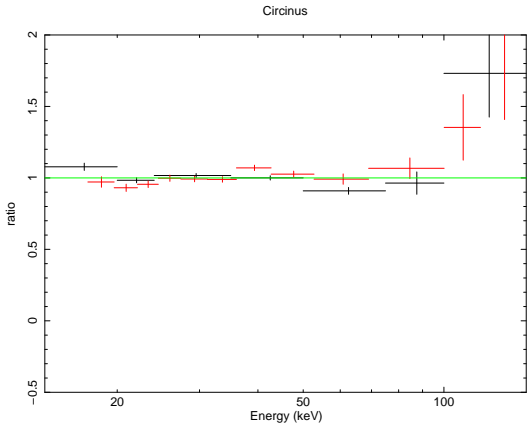}\\
\caption{Best fit model-to-\emph{IBIS/BAT} data (\texttt{pexrav}, R free) ratio for the Circinus Galaxy (see Table~\ref{pex_fit2}).  \emph{IBIS} data are in red while BAT data are in black. 
As can be seen from the plot, both  \emph{IBIS} and \emph{BAT} high energy  data points are not correctly fitted by the model.}
\label{circinus}
\end{figure}

The fact that a number of sources are still badly fitted after substituting the simple power-law
with the \texttt{pexrav} model could have various explanations, among which
a certain degree of flux variability (see Tables~\ref{sim_fit_type1}/\ref{sim_fit_type2})
and/or poor quality data (either related to \emph{IBIS}, \emph{BAT} or both). In addition, 
there may be further spectral complexity like in Cygnus A, where the hot temperature of the surrounding cluster
could have an effect on the source spectrum, even above 10\,keV \citep{Molina:2006}. However, 
a common problem that we find in these 9 cases is a divergence between model and data points at
either low (i.e. around 17-20\,keV) or high energies (i.e. above 100\,keV).
For instance, in NGC 4945 (Figure~\ref{ngc4945}), the first 
\emph{IBIS} point is causing problems to the fit, while in Cygnus A (Figure~\ref{cyga}) the problem
arises due to the last \emph{BAT} spectral point. On the other hand, in the Circinus Galaxy (Figure~\ref{circinus}), 
\emph{IBIS} and \emph{BAT} points track each other perfectly well, but both diverge 
from the model above 100\,keV; given the good agreement between \emph{IBIS} and \emph{BAT} spectra,
this could be a true feature of the source (which is of some scientific interest) 
or could be due to problems related for example to an incorrect background subtraction. 
This is not an uninteresting issue, which we intend to pursue as more data on this source
become available. In order to test the aforementioned problem, we have refitted all 9 objects using the same approach 
described in section~\ref{spec_compl}, but limiting the analysis to the 20--100\,keV 
energy range. 
We find that the resulting fits become acceptable, at least in one of the various cases of R considered; 
the only exception are IGR J2124+5058 and the Circinus Galaxy, where the reduced $\chi^2$ still remains above 1.5.
Inspection of the data-to-model ratios for these two sources indicates that 
the fits are nonetheless acceptable, suggesting that these may just be the 2-3
objects where the reduced $\chi^2$ is above 1.5 just by chance in a sample of 28 objects. 
As a final remark, we note that, in general, the parameter values obtained in this restricted energy range are consistent 
with those obtained in the 17--150\,keV fits.

\begin{table*}
\footnotesize
\centering
\caption{{\bf BAT/IBIS simultaneous fits - \texttt{pexrav} model, type 1 AGN}}
\label{pex_fit1}
\begin{tabular}{lcccccr}
\hline
{\bf Name} & {\bf $\bf \Gamma$} & {\bf E$_{\rm \bf c}$ (keV)} &{\bf R}&{\bf C$_{\rm \bf IBIS/BAT}$} & {\bf $\bf \chi^2$ (d.o.f.)}& {\bf prob.$^{\dagger}$}\\
\hline
MCG+08-11-011&1.51$^{+0.62}_{-0.40}$& $>$38            & $<$3.67&1.05$^{+0.22}_{-0.21}$&9.4 (14)& $>$99.9\%\\
             &1.35$^{+0.23}_{-0.25}$&56$^{+30}_{-15}$  &  0f    &1.05$^{+0.22}_{-0.21}$&9.8 (15)& $>$99.9\% \\
             &1.79$^{+0.21}_{-0.22}$&148$^{+385}_{-67}$&  1f    &1.05$^{+0.22}_{-0.21}$&9.7 (15)& $>$99.9\%  \\
             &1.99$^{+0.15}_{-0.21}$&       $>$137     &  2f    &1.05$^{+0.22}_{-0.21}$&10.6 (15)&$>$99.9\%  \\
\hline
Mrk 6     &1.29$_{-0.38}^{+0.51}$&68$_{-26}^{+276}$&$<$1.06&1.09$_{-0.14}^{+0.15}$& 18.0 (14)& 94\%\\
          &1.29$_{-0.38}^{+0.35}$&68$_{-26}^{+88}$ &0f     &1.09$_{-0.14}^{+0.15}$& 18.0 (15)& 98.6\%\\
          &1.70$_{-0.36}^{+0.28}$&$>$76            &1f     &1.09$_{-0.14}^{+0.15}$& 20.5 (15)& 95.7\%\\
          &1.87$_{-0.33}^{+0.17}$&$>$118           &2f     &1.09$_{-0.14}^{+0.15}$& 22.0 (15)& 92.1\%\\
\hline
NGC 3783     &1.94$^{+0.12}_{-0.37}$&$>$100           &0.82$^{+0.80}_{-1.10}$&1.18$^{+0.20}_{-0.20}$&20.3 (14) & 99.3\%\\
             &1.58$^{+0.15}_{-0.16}$&102$^{+69}_{-31}$&   0f  &1.18$^{+0.20}_{-0.20}$&23.5 (15)& 99.6\%\\
             &1.99$^{+0.07}_{-0.15}$&$>$262           &   1f  &1.18$^{+0.20}_{-0.20}$&20.3 (15)& 99.9\%\\
             &2.04$^{+0.04}_{-0.07}$&$>$942           &   2f  &1.18$^{+0.20}_{-0.20}$&23.2 (15)& 99.6\%\\
\hline
{\bf NGC 4151}&{\bf 1.59$^{+0.09}_{-0.04}$}&{\bf 158$^{+76}_{-13}$}&{\bf $<$0.16}&{\bf 0.93$^{+0.02}_{-0.02}$}&{\bf 24.9 (14)}&{\bf $>$99.9\%}\\
              &{\bf 1.59$^{+0.04}_{-0.04}$}&{\bf 151$^{+23}_{-18}$}&{\bf  0f}&{\bf 0.93$^{+0.02}_{-0.02}$}&{\bf 25.0 (15)}&{\bf $>$99.9\%} \\
              &{\bf 1.91$^{+0.01}_{-0.01}$}&{\bf NC}&{\bf 1f}&{\bf 0.92$^{+0.02}_{-0.02}$}&{\bf 97.5 (15)}&{\bf $>$99.9\%}\\
              &{\bf 1.94$^{+0.01}_{-0.01}$}&{\bf NC} & {\bf 2f} &{\bf 0.92$^{+0.02}_{-0.02}$}&{\bf 278.7 (15)}&{\bf $>$99.9\%}\\
\hline
NGC 4593     &1.48$_{-0.25}^{+0.28}$&100$^{+195}_{-38}$&$<$0.39&1.33$_{-0.08}^{+0.08}$&15.9 (14)& 96\% \\
             &1.48$_{-0.25}^{+0.24}$&100$_{-38}^{+128}$&0f&1.33$_{-0.08}^{+0.09}$& 15.9 (15)& 99\% \\
             &1.87$_{-0.24}^{+0.13}$&$>$157&1f&1.32$_{-0.08}^{+0.08}$& 21.3 (15)&94\% \\
             &1.97$_{-0.17}^{+0.06}$&NC&2f&1.32$_{-0.08}^{+0.09}$& 25.0 (15)& 88\% \\
\hline
MCG-06-30-015&2.55$^{+0.11}_{-1.41}$&$>$94&NC&1.07$^{+0.13}_{-0.12}$&20.4 (14)& 96\%\\
             & 1.44$^{+0.54}_{-0.66}$& 35$^{+45}_{-15}$&0f& 1.06$^{+0.13}_{-0.12}$& 22.6 (15)& 97.8\%\\
             & 1.90$^{+0.52}_{-0.58}$& $>$31&1f& 1.06$^{+0.13}_{-0.12}$& 21.3 (15)& 98.7\%\\
             & 2.19$^{+0.38}_{-0.56}$& $>$42&2f& 1.07$^{+0.13}_{-0.12}$& 20.9 (15)& 98.9\%\\
\hline
4U 1344-60   &1.47$^{+0.31}_{-0.22}$&72$^{+86}_{-20}$&$<$0.58             &1.12$^{+0.06}_{-0.05}$&16.5 (14)& 99.8\%\\
             &1.47$^{+0.21}_{-0.22}$&72$^{+40}_{-20}$&      0f             &1.12$^{+0.06}_{-0.05}$&15.5 (15)&$>$99.9\%\\
             &1.84$^{+0.19}_{-0.21}$&202$^{+745}_{-96}$&    1f             &1.12$^{+0.06}_{-0.05}$&20.8 (15)&99.8\%\\
             &1.99$^{+0.14}_{-0.19}$&$>$181&    2f             &1.12$^{+0.06}_{-0.05}$&23.9 (15)&99.5\%\\
\hline
{\bf IC 4329A}&{\bf 1.64$^{+0.29}_{-0.25}$}&{\bf 101$^{+196}_{-41}$}&{\bf 0.61$^{+0.70}_{-0.50}$}&{\bf 1.13$^{+0.04}_{-0.03}$}&{\bf 23.5 (14)}&{\bf$>$99.9\%}\\
             &{\bf 1.35$^{+0.11}_{-0.11}$}&{\bf 57$^{+11}_{-8}$}& {\bf 0f}&{\bf  1.13$^{+0.04}_{-0.03}$}&{\bf  28.1 (15)}& {\bf $>$99.9\%}\\
             &{\bf 1.76$^{+0.10}_{-0.11}$}&{\bf 143$^{+71}_{-37}$} & {\bf 1f}& {\bf 1.13$^{+0.04}_{-0.03}$}& {\bf 23.9 (15)}&{\bf  $>$99.9\%}\\
             &{\bf 1.95$^{+0.10}_{-0.10}$}&{\bf 335$^{+567}_{-134}$}&{\bf  2f}& {\bf 1.13$^{+0.04}_{-0.03}$}& {\bf 27.2 (15)}&{\bf  $>$99.9\%}\\
\hline
GRS 1734-292 &1.98$^{+0.30}_{-0.27}$&$<$60&$<$1.41          &1.19$^{+0.05}_{-0.05}$&17.5 (14)&99.9\%\\             
             &1.83$^{+0.14}_{-0.14}$&89$^{+38}_{-22}$&    0f             &1.19$^{+0.05}_{-0.05}$&18.4 (15) &99.9\%\\
             &2.15$^{+0.13}_{-0.13}$&$>$161&  1f             &1.19$^{+0.05}_{-0.05}$&18.4 (15)& 99.9\%\\
             &2.30$^{+0.05}_{-0.12}$&NC         &  2f             &1.19$^{+0.05}_{-0.05}$&20.8 (15)&99.9\%\\
\hline
{\bf IGR J17488-3253}&{\bf 1.81$^{+0.06}_{-0.06}$}& {\bf NC} &{\bf $<$0.07 }&{\bf 2.11$^{+0.28}_{-0.23}$}&{\bf 44.0 (14)}& {\bf...}\\               
&{\bf 1.81$^{+0.06}_{-0.06}$}&{\bf NC}&{\bf 0f}&{\bf 2.11$^{+0.28}_{-0.23}$}&{\bf 44.0 (15)}&... \\
               &{\bf 1.81$^{+0.07}_{-0.05}$}&{\bf NC}&{\bf 1f}&{\bf 2.11$^{+0.28}_{-0.23}$}&{\bf 75.1 (15)}&...\\
               &{\bf 1.81$^{+0.08}_{-0.05}$}&{\bf NC}&{\bf 2f}&{\bf 2.11$^{+0.28}_{-0.24}$}&{\bf 92.4 (15)}&...\\
\hline
{\bf IGR J18027-1455}&{\bf 1.83$_{-0.12}^{+0.08}$}&{\bf NC}&{\bf $<$0.15}&{\bf 2.17$_{-0.28}^{+0.32}$}&{\bf 28.3 (14)} & ... \\
&{\bf 1.81$_{-0.10}^{+0.10}$}&{\bf NC}&{\bf 0f}&{\bf 2.14$_{-0.26}^{+0.34}$}&{\bf 28.0 (15)}& ... \\
                &{\bf 1.81$_{-0.09}^{+0.12}$}&{\bf NC}&{\bf 1f}&{\bf 2.14$_{-0.28}^{+0.32}$}&{\bf 41.7 (15)}& ... \\
                &{\bf 1.81$_{-0.08}^{+0.13}$}&{\bf NC}&{\bf 2f}&{\bf 2.14$_{-0.28}^{+0.31}$}&{\bf 49.1 (15)}& ... \\
\hline
4C 74.26        &1.93$_{-0.70}^{+0.35}$&$>$62&$<$6.65&1.34$_{-0.35}^{+0.35}$& 21.3 (14)& 85\%\\
                &1.69$_{-0.49}^{+0.43}$&$>40$&0f&1.34$_{-0.35}^{+0.35}$& 21.6 (15)& 95\% \\
                &2.05$_{-0.45}^{+0.23}$&$>74$&1f&1.34$_{-0.35}^{+0.35}$& 21.3 (15)& 95\%\\
                &2.20$_{-0.41}^{+0.11}$&NC&2f&1.34$_{-0.35}^{+0.35}$& 21.4 (15)& 95\%\\
\hline

\multicolumn{7}{l}{$^{\dagger}$: F-test probability obtained from the comparison between the 
simple power-law fits and the}\\
\multicolumn{7}{l}{current model.}\\
\multicolumn{7}{l}{{\it Note:} sources with $\Delta$$\chi^2$$>$1.5 are highlighted in bold typeface.}\\
\end{tabular}
\end{table*}

\begin{table*}
\footnotesize
\centering
\caption{{\bf BAT/IBIS simultaneous fits - \texttt{pexrav} model, type 1 AGN - continued}}
\label{pex_fit1_cont}
\begin{tabular}{lcccccr}
\hline
{\bf Name} & {\bf $\bf \Gamma$} & {\bf E$_{\rm \bf c}$ (keV)} &{\bf R}&{\bf C$_{\rm \bf IBIS/BAT}$} & {\bf $\bf \chi^2$ (d.o.f.)}& {\bf prob.$^{\dagger}$}\\

\hline
{\bf IGR J21247+5058}&{\bf 1.72$^{+0.14}_{-0.10}$}&{\bf 155$^{+157}_{-41}$}&{\bf $<$0.23}&{\bf 1.02$^{+0.03}_{-0.02}$}&{\bf 27.7 (14)}& {\bf 98.8\%}\\
               &{\bf 1.72$^{+0.09}_{-0.10}$}&{\bf 155$^{+84}_{-41}$}&{\bf 0f}&{\bf 1.02$^{+0.03}_{-0.03}$}&{\bf 27.7 (15)}&{\bf 99.7\%}\\
               &{\bf 2.03$^{+0.02}_{-0.06}$}&{\bf NC}&{\bf 1f}&{\bf 1.01$^{+0.03}_{-0.02}$}&{\bf 42.7 (15)}&{\bf 89.9\%}\\
               &{\bf 2.06$^{+0.02}_{-0.04}$}&{\bf NC}&{\bf 2f}&{\bf 1.01$^{+0.03}_{-0.02}$}&{\bf 67.0 (15)}&{\bf ...}\\

\hline
MR 2251-178 &1.37$_{-0.30}^{+0.67}$&$>$42&$<2.26$&0.96$_{-0.20}^{+0.20}$& 10.2 (14)&99.9\% \\
            &1.37$_{-0.30}^{+0.28}$&62$_{-20}^{+49}$&0f&0.96$_{-0.20}^{+0.20}$&10.2 (15) & $>$99\%\\
            &1.78$_{-0.28}^{+0.25}$&$>$80&1f&0.95$_{-0.20}^{+0.20}$& 11.4 (15) & 99.9\%\\
            &1.96$_{-0.26}^{+0.15}$&$>$128&2f&0.95$_{-0.20}^{+0.20}$&12.5 (15) & 99.9\%\\
          
\hline
\multicolumn{7}{l}{$^{\dagger}$: F-test probability obtained from the comparison between the simple power-law fits and the}\\
\multicolumn{7}{l}{current model.}\\
\multicolumn{7}{l}{{\it Note:} sources with $\Delta$$\chi^2$$>$1.5 are highlighted in bold typeface.}\\
\end{tabular}
\end{table*}

\begin{table*}
\footnotesize
\centering
\caption{{\bf BAT/IBIS simultaneous fits - \texttt{pexrav} model, type 2 AGN}}
\label{pex_fit2}
\begin{tabular}{lcccccr}
\hline
{\bf Name} & {\bf $\bf \Gamma$} & {\bf E$_{\rm \bf c}$ (keV)} &{\bf R}&{\bf C$_{\rm \bf IBIS/BAT}$} & {\bf $\bf \chi^2$ (d.o.f.)}& {\bf prob$^{\dagger}$}\\
\hline
MCG-05-23-16&1.65$^{+0.40}_{-0.38}$&72$^{+152}_{-31}$&$<$2.82&1.04$^{+0.08}_{-0.08}$&20.6 (14)&$>$99.9\%\\
&1.37$^{+0.14}_{-0.15}$&47$^{+11}_{-8}$&    0f   &1.04$^{+0.08}_{-0.08}$&22.1 (15)&$>$99.9\%\\
            &1.73$^{+0.13}_{-0.14}$&85$^{+35}_{-20}$&   1f   &1.04$^{+0.08}_{-0.08}$&20.8 (15)&$>$99.9\%\\
            &1.93$^{+0.12}_{-0.13}$&139$^{+102}_{-44}$& 2f   &1.04$^{+0.08}_{-0.08}$&22.1 (15)&$>$99.9\%\\
\hline
NGC 3281    &1.26$^{+0.67}_{-0.43}$&43$^{+123}_{-16}$&$<$3.71 &0.78$^{+0.17}_{-0.16}$&13.3 (14)& 99.9\%\\
   &1.24$^{+0.32}_{-0.47}$&42$^{+23}_{-15}$&    0f   &0.78$^{+0.17}_{-0.16}$&13.2 (15)&99.9\%\\
            &1.57$^{+0.31}_{-0.45}$&68$^{+70}_{-29}$&    1f   &0.78$^{+0.17}_{-0.16}$&14.3 (15)&99.9\%\\
            &1.74$^{+0.30}_{-0.43}$&96$^{+180}_{-47}$&   2f   &0.78$^{+0.17}_{-0.16}$&15.1 (15)&99.9\%\\
\hline
NGC 4388    &1.60$^{+0.11}_{-0.08}$&202$^{+198}_{-154}$&$<$0.23  &1.52$^{+0.04}_{-0.04}$&13.4 (14)& 99.9\%\\            
    &1.60$^{+0.08}_{-0.08}$&202$^{+111}_{-54}$&    0f    &1.52$^{+0.04}_{-0.04}$&13.4 (15)& 99.9\%\\
            &1.85$^{+0.02}_{-0.05}$&     NC        &    1f    &1.52$^{+0.04}_{-0.04}$&27.2 (15)&95.9\%\\
            &1.88$^{+0.02}_{-0.02}$&     NC        &    2f    &1.51$^{+0.04}_{-0.04}$&52.2 (15)&...\\
\hline
NGC 4507    &1.53$^{+0.37}_{-0.34}$&109$^{+647}_{-53}$&$<$2.37&1.33$^{+0.07}_{-0.07}$&14.6 (14)&$>$99.9\%\\            
    &1.34$^{+0.14}_{-0.19}$&72$^{+23}_{-18}$&     0f  &1.34$^{+0.07}_{-0.07}$&15.8 (15)&$>$99.9\%\\
            &1.63$^{+0.15}_{-0.16}$&145$^{+134}_{-50}$&   1f  &1.33$^{+0.07}_{-0.07}$&14.8 (15)&$>$99.9\%\\
            &1.79$^{+0.14}_{-0.15}$&$>$150& 2f   &1.33$^{+0.07}_{-0.07}$&16.1 (15)&$>$99.9\%\\
\hline
{\bf NGC 4945}&{\bf 1.37$^{+0.21}_{-0.18}$}&{\bf 101$^{+262}_{-55}$}&{\bf$<$2.18}&{\bf 1.56$^{+0.04}_{-0.03}$}&{\bf 27.5 (14)}&{\bf $>$99.9\%}\\            
&{\bf 1.37$^{+0.10}_{-0.10}$}&{\bf 101$^{+26}_{-18}$}&{\bf 0f}&{\bf 1.56$^{+0.04}_{-0.03}$}&{\bf 27.5 (15)}&{\bf$>$99.9\%}\\
            &{\bf 1.59$^{+0.10}_{-0.10}$}&{\bf 218$^{+136}_{-63}$} &{\bf 1f}&{\bf 1.55$^{+0.03}_{-0.03}$}&{\bf 26.6 (15)}&{\bf$>$99.9\%}\\ 
            &{\bf 1.71$^{+0.09}_{-0.10}$}&{\bf 445$^{+835}_{-186}$}&{\bf 2f}&{\bf 1.55$^{+0.03}_{-0.03}$}&{\bf 29.6 (15)}&{\bf$>$99.9\%}\\
\hline
{\bf IGR J13091+1137}&{\bf 1.09$^{+0.41}_{-0.63}$}&{\bf 68$^{+119}_{-24}$}&{\bf $<$1.02}&{\bf 1.14$^{+0.21}_{-0.20}$}&{\bf 29.8 (14)}&{\bf 93.9\%}\\              
&{\bf 1.09$^{+0.36}_{-0.66}$}&{\bf 67$^{+80}_{-23}$}&{\bf 0f}&{\bf 1.14$^{+0.21}_{-0.20}$}&{\bf 29.8 (15)}&{\bf 93.9\%}\\
               &{\bf 1.27$^{+0.47}_{-0.48}$}&{\bf $>$61}&{\bf 1f}&{\bf 1.13$^{+0.21}_{-0.20}$}&{\bf 32.1 (15)}&{\bf 88\%}\\
               &{\bf 1.47$^{+0.38}_{-0.50}$}&{\bf$>$85}&{\bf 2f}&{\bf 1.13$^{+0.21}_{-0.20}$}&{\bf 33.5 (15)}&{\bf 82\%}\\
\hline
{\bf Cen A} &{\bf 1.89$^{+0.01}_{-0.01}$}&{\bf NC}&{\bf $<$0.007}&{\bf 0.71$^{+0.01}_{-0.01}$}&{\bf 75.6 (14)}&... \\
 &{\bf 1.89$^{+0.01}_{-0.01}$}&{\bf NC}&{\bf 0f}&{\bf 0.71$^{+0.01}_{-0.01}$}&{\bf 75.6 (15)}& ...\\
            &{\bf 1.94$^{+0.01}_{-0.01}$}&{\bf NC}&{\bf 1f}&{\bf 0.70$^{+0.01}_{-0.01}$}&{\bf 841.1 (15)}&...\\
            &{\bf 1.96$^{+0.01}_{-0.01}$}&{\bf NC}&{\bf 2f}&{\bf 0.70$^{+0.01}_{-0.01}$}&{\bf 1649.9 (15)}&...\\
\hline
{\bf Circinus}&{\bf 1.65$^{+0.28}_{-0.20}$}&{\bf 41$^{+19}_{-10}$}&{\bf $<$1.42}&{\bf 1.15$^{+0.02}_{-0.02}$}&{\bf 64.1 (14)}&{\bf$>$99.9\%}\\
&{\bf 1.50$^{+0.11}_{-0.13}$}&{\bf 34$^{+4}_{-4}$}&{\bf 0f}&{\bf 1.15$^{+0.02}_{-0.02}$}&{\bf 65.0 (15)}&{\bf$>$99.9\%}\\
            &{\bf 1.81$^{+0.11}_{-0.11}$}&{\bf 51$^{+9}_{-7}$}  &{\bf 1f}&{\bf 1.15$^{+0.02}_{-0.02}$}&{\bf 65.3 (15)}&{\bf$>$99.9\%}\\
            &{\bf 2.00$^{+0.11}_{-0.11}$}&{\bf 68$^{+15}_{-11}$}&{\bf 2f}&{\bf 1.15$^{+0.02}_{-0.02}$}&{\bf 69.0 (15)}&{\bf$>$99.9\%}\\
\hline
NGC 5506    &2.01$^{+0.26}_{-0.34}$&  $>$88      &8.90$^{+136.91}_{-6.12}$&1.14$^{+0.21}_{-0.21}$&15.0 (14)&$>$99.9\%\\            
    & 1.04$^{+0.13}_{-0.19}$& 36$^{+5}_{-6}$  & 0f& 1.15$^{+0.21}_{-0.21}$& 37.7 (15)&$>$99.9\%\\
            & 1.39$^{+0.15}_{-0.15}$& 54$^{+14}_{-10}$& 1f& 1.14$^{+0.21}_{-0.21}$& 23.6 (15)&$>$99.9\%\\
            & 1.60$^{+0.14}_{-0.14}$& 76$^{+27}_{-16}$& 2f& 1.14$^{+0.21}_{-0.21}$& 19.2 (15)&$>$99.9\%\\
\hline
\multicolumn{7}{l}{$^{\dagger}$: F-test probability obtained from the comparison between the simple power-law fits and the}\\
\multicolumn{7}{l}{current model.}\\
\multicolumn{7}{l}{{\it Note:} sources with $\Delta$$\chi^2$$>$1.5 are highlighted in bold typeface.}\\
\end{tabular}
\end{table*}

\begin{table*}
\footnotesize
\centering
\caption{{\bf BAT/IBIS simultaneous fits - \texttt{pexrav} model, type 2 AGN - continued}}
\label{pex_fit2_cont}
\begin{tabular}{lcccccr}
\hline
{\bf Name} & {\bf $\bf \Gamma$} & {\bf E$_{\rm \bf c}$ (keV)} &{\bf R}&{\bf C$_{\rm \bf IBIS/BAT}$} & {\bf $\bf \chi^2$ (d.o.f.)}& {\bf prob$^{\dagger}$}\\
\hline
IGR J14515-5542&1.36$_{-0.30}^{+0.52}$&$>$74&$<$1.05&0.69$_{-0.11}^{+0.12}$& 23.67 (14)& 37\%\\
               &1.40$_{-0.59}^{+0.48}$&$>$46&0f&0.69$_{-0.11}^{+0.12}$& 23.66 (15)& 67\% \\
               &1.77$_{-0.58}^{+0.23}$&$>81$&1f&0.68$_{-0.11}^{+0.12}$& 26.29 (15)& ... \\
               &1.84$_{-0.51}^{+0.16}$&$>$150&2f&0.68$_{-0.11}^{+0.12}$& 27.56 (15)&  ... \\
\hline
IC 4518A    &0.65$_{-1.35}^{+1.29}$&20$_{-3}^{+64}$&$<$31&1.10$_{-0.16}^{+0.20}$& 11.74 (14)& 99.9\%\\
            &0.61$_{-1.48}^{+0.82}$&20$_{-9}^{+13}$&0f&1.10$_{-0.16}^{+0.20}$& 11.70 (15)&99.9\% \\
            &0.85$_{-1.23}^{+1.00}$&24$_{-11}^{+40}$&1f&1.10$_{-0.16}^{+0.20}$& 12.67 (15)&99.8\% \\
            &1.11$_{-1.20}^{+0.96}$&30$_{-15}^{+75}$&2f&1.10$_{-0.16}^{+0.20}$& 13.16 (15)& 99.8\%\\
\hline
NGC 6300      &2.18$^{+0.11}_{-0.78}$&     $>$518   &$>$0.49 &1.11$^{+0.09}_{-0.09}$&12.27 (14)&$>$99.9\%\\
     &1.25$^{+0.28}_{-0.42}$&44$^{+21}_{-15}$&  0f         &1.10$^{+0.09}_{-0.09}$&16.66 (15)&99.98\%\\
              &1.57$^{+0.28}_{-0.38}$&71$^{+67}_{-29}$&  1f         &1.10$^{+0.09}_{-0.09}$&14.04 (15)&$>$99.9\%\\
              &1.67$^{+0.34}_{-0.29}$&88$^{+213}_{-34}$& 2f         &1.10$^{+0.09}_{-0.09}$&13.22 (15)&$>$99.9\%\\              
\hline
ESO 103-G35  &1.81$^{+0.50}_{-0.84}$&$>$31&$<$34.23&0.82$^{+0.23}_{-0.23}$&14.31 (14)& $>$99.9\%\\
   &1.17$^{+0.25}_{-0.37}$&38$^{+13}_{-11}$&  0f     &0.82$^{+0.23}_{-0.23}$&16.52 (15)&$>$99.9\%\\
              &1.51$^{+0.24}_{-0.35}$&61$^{+36}_{-22}$&  1f     &0.82$^{+0.23}_{-0.23}$&14.80 (15)&$>$99.9\%\\
              &1.63$^{+0.31}_{-0.25}$&74$^{+95}_{-24}$&  2f     &0.82$^{+0.23}_{-0.23}$&14.44 (15)&$>$99.9\%\\
             
\hline
{\bf Cyg A}   &{\bf 2.01$^{+0.04}_{-0.04}$}&{\bf NC}&{\bf $<$0.09}&{\bf 1.03$^{+0.04}_{-0.04}$}&{\bf 33.85 (14)}&...\\         
   &{\bf 2.01$^{+0.04}_{-0.04}$}&{\bf NC}&{\bf 0f}&{\bf 1.03$^{+0.04}_{-0.04}$}&{\bf 33.85 (15)}&...\\
              &{\bf 2.04$^{+0.04}_{-0.04}$}&{\bf NC}&{\bf 1f}&{\bf 1.03$^{+0.04}_{-0.04}$}&{\bf 63.52 (15)}&...\\
              &{\bf 2.06$^{+0.04}_{-0.04}$}&{\bf NC}&{\bf 2f}&{\bf 1.02$^{+0.04}_{-0.04}$}&{\bf 86.16 (15)}&...\\              
\hline
\multicolumn{7}{l}{$^{\dagger}$: F-test probability obtained from the comparison between the simple power-law fits and the}\\
\multicolumn{7}{l}{current model.}\\
\multicolumn{7}{l}{{\it Note:} sources with $\Delta$$\chi^2$$>$1.5 are highlighted in bold typeface.}\\
\end{tabular}
\end{table*}

\section{The high energy continuum and its cut-off energy}\label{cut_off}

\begin{figure}
\centering
\includegraphics[width=1.0\linewidth]{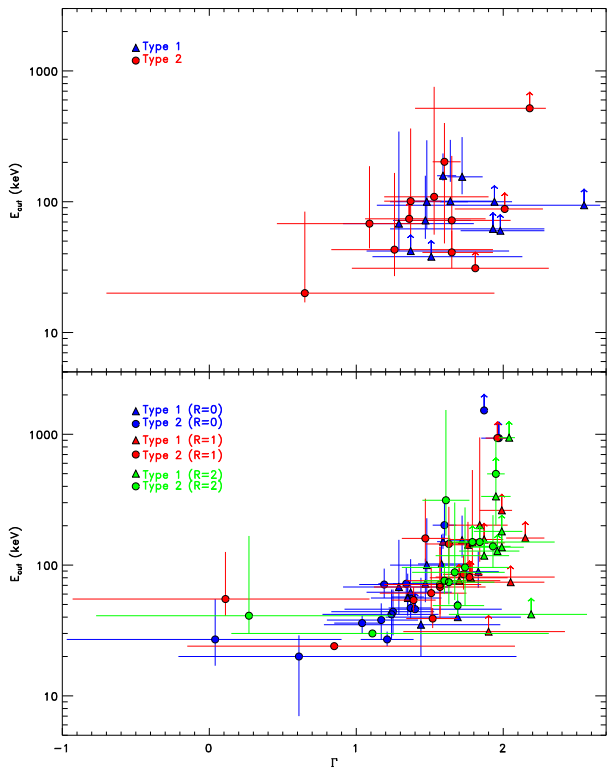}
\caption{High energy cut-off \emph{vs.} photon index for
those AGN which display spectral complexity. Data are from Table 8,9,10,11. {\it Upper panel}: parameter space
relative to fits where R was left as a free variable; blue triangles represent type 1 sources, red circles represent type 2 objects.
Upper limits on the cut-off energy are represented by arrows.
{\it Lower panel}: parameter space relative to fits where R was assumed to be 0, 1 or 2. 
Triangles represent type 1 sources while circles indicate type 2 objects; different colours refer to different values of R (R=0 in blue,
R=1 in red, R=2 in green). Upper limits on the cut-off energy are represented by arrows.}
\label{g_ecut}
\end{figure}

In Figure~\ref{g_ecut}, we show the high energy cut-off (E$_{\rm cut}$) \emph{vs.} the
photon index for the AGN where these two quantities could be measured
(data are either from Table~\ref{pex_fit1}/ \ref{pex_fit1_cont} and \ref{pex_fit2}/\ref{pex_fit2_cont}); 
{\bf in the figure upper panel, we show the parameter space obtained when leaving
R as a free parameter in the fits, while in the lower panel, we show the parameter space
obtained when R is allow to have only the values of 0, 1 and 2.}
Different colouring and symbols reflect the different R values assumed 
and the different AGN types analysed. 
One thing which is immediately evident is that the two figures are fully compatible,
suggesting that the choice of fixing R is acceptable if one wants to have more
precise information on a large sample of sources. The second evident thing
is the \emph{banana} shape of the space covered by the parameters explored in both plots;
this is a clear sign of the inter-relation between the two  
analysed quantities, in the sense that flatter spectra provide 
lower cut-off energies than steeper ones. This again underlines the difficulties
of extracting spectral information from the analysis of narrow band data.
From the figure, it is also evident that the sample covers 
quite a large range in photon indices as well as cut-off energies; 
however, as can be seen in the lower panel of Figure~\ref{g_ecut}, 
the spread is due only to a small number of objects 
($\lesssim$20\% of the sample) while 
the majority of our AGN lies within well defined boundaries of photon index 
(1$\le$$\Gamma$$\le$2) and cut-off energy (30$\le$E$_{\rm cut}$$\le$300\,keV). 
In fact, the average power-law photon index and high energy cut-off for
all AGN where both could be measured are
$\Gamma$ = 1.39, 1.62, 1.75 ($\sigma$ = 0.39 in each case)
and E$_{\rm cut}$ = 215, 204, 205 \,keV ($\sigma$ = 368, 274, 295\,keV) assuming R = 0, 1, and 2
respectively; note that lower limits on the high energy cut-off have been considered.

Our results are in line with \emph{BeppoSAX} observations which 
indicate a spread in E$_{\rm cut}$ from about 50\,keV up to around 300\,keV \citep{Perola:2002, Dadina:2008};
they also agree with the results of \citet{beckmann09},
but are barely consistent with those obtained by \citet{ricci11}, who locate the cut-off energy above 200-300\,keV.

The values of $\Gamma$ and E$_{\rm cut}$ are linked to the Comptonising
hot plasma temperature kT$_{\rm e}$ and optical depth $\tau$, according to the
relation discussed by \citet{Petrucci:2001} and also analysed by \citet{Molina:2009}.

In particular, the plasma temperature kT$_{\rm e}$ is estimated as kT$_{\rm e}$ = E$_{\rm cut}$/2
if $\tau$$\lesssim$1 and kT$_{\rm e}$ = E$_{\rm cut}$/3 if $\tau$$\gg$1. 
The most likely range of E$_{\rm cut}$ estimated in the present sample (30-300\,keV),
indicates a range of plasma temperatures from 10 to 150\,keV (or
1.16$\times$10$^8$-1.74$\times$10$^9$\,K). The equation (solved for both low and high values of
$\tau$ and E$_{\rm cut}$ and assuming our average value of $\Gamma$=1.65 in the case of R=1) has therefore
acceptable solutions for $\tau$ in the range $\sim$1 to $\sim$6. These results are in
good agreement with what was previously found for a small sample
of Seyfert 1 galaxies studied by \citet{Petrucci:2001} and \citet{Molina:2009} and indicate
that the Comptonising hot plasma has a typical temperature of 80$\pm$70\,keV and is
mildly thick ($\tau$$<$7). Available models should be able to explain and
cover the observed range of values.

As a final remark, we also note that synthesis model of the cosmic diffuse background (CXB) often assume as an upper
limit to the cut-off energy a value of $\sim$200\,keV; this choice is basically driven by the intensity and the shape
of the CXB spectrum above the peak, which cannot be exceeded; even a value of 300\,keV has difficulties in accommodating 
all available observations and CXB measurements \citep{Gilli:2007}. Thus our analysis agrees and supports this 
CXB model assumption.

\section{Summary and Conclusions}

In this work we present the hard X-ray properties of a complete sample
of AGN using both \emph{INTEGRAL/IBIS} and for 
the majority of the sources also \emph{Swift/BAT} data. In particular,
the main objective of this work is to analyse the cross-calibration between 
\emph{IBIS} and \emph{BAT}, study source variability and constraining some
important spectral parameters such as the high energy continuum and its
cut-off energy. The main results found in this paper can be summarised as follows:

\begin{itemize}
 \item[-] We find that out of 80 sources for which both \emph{BAT} and
 \emph{IBIS} spectra are available, 11 show signs of flux variability. This result
 suggests that flux variability is not uncommon at high energies, although it is
 not very dramatic (see Section~\ref{var}). Spectral variability is instead quite rare,
 being found only in one source (i.e. 3C 273).
 \item[-] There is general good agreement between \emph{INTEGRAL/IBIS} and
  \emph{Swift/BAT} data, although we find a systematic difference in the normalisation
  between the two instruments of about 22\%. This is also reflected in the combined fits
  discussed in Section~\ref{cross_calib}, where we find that the average cross-calibration
  constant between \emph{IBIS/BAT} is 1.22.
 \item[-] We find, from the combined fits reported in Section~\ref{cross_calib}, that the
 average photon index for our sample is around 2, with quite a narrow spread. We also find that
 type 1 and type 2 sources have very similar average photon indices, a clear indication
 that they are powered by the same mechanism.
 \item[-] Another interesting result is that a simple power-law is not always
 the most appropriate model to describe high energy data. Indeed, 35\% of the sources
 show signs of spectral complexity (see Section~\ref{spec_compl}) and therefore
 have to be fitted with a model that takes into account a high energy cut-off 
 and a reflection component. 
 \item[-] This more complex model yields a fit improvement in 75\% of the sources considered,
 thus providing evidence for the presence of a high energy component.
 \item[-] Although it is quite hard to constrain at once the high energy continuum, its
 high energy cut-off and the reflection fraction, it is possible to put constraints on at 
 least two parameters while the other is fixed. By fixing the reflection fraction to be 0, 1 and 2,
 we have been able to determine the parameter space defined by the photon index and the high energy cut-off.
 \item[-] Our sample covers quite a large range 
 in photon indices as well as cut-off energies; however, the spread is due only to a small number of objects 
 while the majority of the AGN lie within well defined boundaries of photon index 
 (1$\le$$\Gamma$$\le$2) and cut-off energy (30$\le$E$_{\rm cut}$$\le$300\,keV).
 \item[-] The values of $\Gamma$ and E$_{\rm cut}$ are linked to the Comptonising
hot plasma temperature kT$_{\rm e}$ and optical depth $\tau$; from our analysis
we find that, for the sources in our sample, the plasma has a typical temperature of (80$\pm$70)\,keV and is
mildly thick ($\tau$$<$7). 
\end{itemize}

\section*{Acknowledgements}
The authors acknowledge financial support from ASI under contracts ASI/033/10/0 and ASI/INAF I/009/10/0.


\begin{thebibliography}{99}
\bibitem[\protect\citeauthoryear{Barthelmy et al.}{2005}]{barthelmy05} Barthelmy S.~D., Barbier L.~M.,  
Cummings J.~R. et al. 2005, Space Science Reviews, 120, 143
\bibitem[\protect\citeauthoryear{Baumgartner et al.}{2010}]{baumgartner10} Baumgartner W.~H., Tueller J., 
Markwardt C., Skinner G., 2010, in  Bulletin of the American Astronomical Society, Vol.~42, 
AAS/High Energy Astrophysics Division \#11, p. 675
\bibitem[\protect\citeauthoryear{Beckmann et al.}{2006}]{beckmann06} Beckmann V., 
Soldi S., Shrader C.~R., Gehrels N., Produit N., 2006, ApJ, 652, 126
\bibitem[\protect\citeauthoryear{Beckmann et al.}{2007}]{Beckmann:2007} Beckmann V., 
Barthelmy S.~D., Courvoisier T., Gehrels N., Soldi S., Tueller J., Wendt G., 2007, A\&A, 475, 827
\bibitem[\protect\citeauthoryear{Beckmann et al.}{2009}]{beckmann09} Beckmann V., Soldi S.,
Ricci C. et al.  2009, A\&A, 505, 417
\bibitem[\protect\citeauthoryear{Bianchi et al.}{2009}]{Bianchi:2009} Bianchi S., Guainazzi M., 
Matt G., Fonseca Bonilla N., Ponti G., 2009, A\&A, 495, 421
\bibitem[\protect\citeauthoryear{Bird et al.}{2007}]{Bird:2007} Bird A.~J., Malizia A., Bazzano A. 
et al., 2007, ApJS, 170, 175
\bibitem[\protect\citeauthoryear{Bird et al.}{2010}]{Bird:2010} Bird A.~J., Bazzano A., Bassani L.
et al., 2010, ApJS, 186, 1
\bibitem[\protect\citeauthoryear{Burlon et al.}{2011}]{burlon11} Burlon D., Ajello M., Greiner J., 
Comastri A., Merloni A., Gehrels N., 2011, ApJ, 728, 58
\bibitem[\protect\citeauthoryear{Corral et al.}{2011}]{corral11} Corral A., Della Ceca R., Caccianiga A., 
Severgnini P., Brunner H., Carrera F.~J., Page M.~J., Schwope A.~D., 2011, A\&A, 530, A42
\bibitem[\protect\citeauthoryear{Dadina}{2007}]{Dadina:2007} Dadina M., 2007, A\&A, 461, 1209
\bibitem[\protect\citeauthoryear{Dadina}{2008}]{Dadina:2008} Dadina M., 2008, A\&A, 485, 417
2012, MNRAS, 420, 2087
\bibitem[\protect\citeauthoryear{Deluit \& Courvoisier}{2003}]{deluit03} Deluit S., Courvoisier T.~J.-L., 
2003, A\&A, 399, 77
\bibitem[\protect\citeauthoryear{Elitzur \& Shlosman}{2006}]{Elitzur:2006} Elitzur M. and Shlosman I., 
2006, ApJL, 648, 101
\bibitem[\protect\citeauthoryear{Elitzur}{2012}]{elitzur12} Elitzur M., 2012, ApJL, 747, 33
\bibitem[\protect\citeauthoryear{Gehrels et al.}{2004}]{gehrels04} Gehrels N., Chincarini G., Giommi P. et al.,
 2004, ApJ, 611, 1005
\bibitem[\protect\citeauthoryear{Gilli et al.}{2007}]{Gilli:2007} Gilli R., Comastri A., Hasinger G., 2007, A\&A, 463, 79
\bibitem[\protect\citeauthoryear{Goldwurm et al.}{2003}]{Goldwurm:2003} Goldwurm A., David P., Foschini L. et al.,
2003, A\&A, 411, L223
\bibitem[\protect\citeauthoryear{Lebrun et al.}{2003}]{Lebrun:2003} Lebrun F., Leray J.~P.,  Lavocat P. et al.,
2003, A\&A, 411, L141
\bibitem[\protect\citeauthoryear{Malizia et al.}{2009}]{Malizia:2009} Malizia A., Stephen J.~B., Bassani L., 
Bird A.~J., Panessa F., Ubertini P., 2009, MNRAS, 399, 944
\bibitem[\protect\citeauthoryear{Molina et al.}{2006}]{Molina:2006} Molina M., Malizia A., Bassani L. et al.,
 2006, MNRAS, 371, 821
\bibitem[\protect\citeauthoryear{Molina et al.}{2007}]{Molina:2007} Molina M., Giroletti M., Malizia A. et al.,
 2007, MNRAS, 382, 937
\bibitem[\protect\citeauthoryear{Molina et al.}{2009}]{Molina:2009} Molina M., Bassani L., Malizia A. et al.,
 2009, MNRAS, 399, 1293
\bibitem[\protect\citeauthoryear{Perola et al.}{2002}]{Perola:2002} Perola G.~C., Matt G., Cappi M., Fiore F., Guainazzi M., 
Maraschi L., Petrucci P.~O., Piro L., 2002, A\&A, 389, 802
\bibitem[\protect\citeauthoryear{Petrucci et al.}{2001}]{Petrucci:2001} Petrucci P.~O., Haardt F., Maraschi, L. 
et al., 2001, ApJ, 556, 716
\bibitem[\protect\citeauthoryear{Ricci et al.}{2011}]{ricci11} Ricci C., Walter R., Courvoisier T.~J.-L., Paltani S., 2011, A\&A, 
532, A102
\bibitem[\protect\citeauthoryear{Tazaki et al.}{2010}]{tazaki10} Tazaki F., Ueda Y., Ishino Y., Eguchi S., Isobe N., 
Terashima Y., Mushotzky R.~F., 2010, ApJ, 721, 1340
\bibitem[\protect\citeauthoryear{Ubertini et al.}{2003}]{Ubertini:2003} Ubertini P., Lebrun F., Di Cocco G. et al.,
2003, A\&A, 411, L131
\bibitem[\protect\citeauthoryear{Urry \& Padovani}{1995}]{Urry:1995} Urry C.~M. \& Padovani P., 1995, PASP, 107, 803
\bibitem[\protect\citeauthoryear{Winkler et al.}{2003}]{winkler03} Winkler C., Courvoisier T.~J.-L., Di Cocco G. et al.
2003, A\&A, 411, L1
\bibitem[\protect\citeauthoryear{Yang et al.}{2009}]{yang09} Yang Y., Wilson A.~S., Matt G., Terashima Y., 
Greenhill L.~J., 2009, ApJ, 691, 131
\bibitem[\protect\citeauthoryear{Zdziarski et al.}{1995}]{zdziarski95} Zdziarski A.~A., Johnson W.~N., Done C., Smith D., 
McNaron-Brown K., 1995, ApJL 438, L63
\end{thebibliography}

\label{lastpage}
\end{document}